\documentclass[printer]{aa} 
\usepackage{graphicx}
\usepackage{txfonts}
\begin{document}
   \title{The VIMOS VLT Deep Survey         
         \thanks{based on data
         obtained with the European Southern Observatory Very Large
         Telescope, Paranal, Chile, program 070.A-9007(A), and on data
         obtained at the Canada-France-Hawaii Telescope, operated by
         the CNRS of France, CNRC in Canada and the University of Hawaii}
}

   \subtitle{First epoch VVDS-Deep survey: 11564 spectra with 
$17.5 \leq I_{AB}\leq24$, and the redshift distribution over $0\leq z \leq5$
}

   \author{
O. Le F\`evre \inst{1}, 
G. Vettolani \inst{2},
B. Garilli \inst{3}, 
L. Tresse  \inst{1},
D. Bottini \inst{3}, 
V. Le Brun \inst{1}, 
D. Maccagni \inst{3}, 
J.P. Picat \inst{4},  
R. Scaramella \inst{2}, 
M. Scodeggio \inst{3}, 
A. Zanichelli \inst{2}, 
C. Adami \inst{1}, 
M. Arnaboldi \inst{5},
S. Arnouts \inst{1},
S. Bardelli \inst{8},
M. Bolzonella \inst{9},
A. Cappi \inst{8}, 
S. Charlot \inst{6,11}, 
P. Ciliegi\inst{8} , 
T. Contini \inst{4},
S. Foucaud \inst{3},  
P. Franzetti \inst{3},
I. Gavignaud \inst{4,12}, 
L. Guzzo \inst{10}, 
O. Ilbert \inst{1,8}, 
A. Iovino \inst{10}, 
H.J. McCracken \inst{6}, 
B. Marano \inst{9}, 
C. Marinoni \inst{1,10},
G. Mathez \inst{4}, 
A. Mazure \inst{1},
B. Meneux \inst{1},
R. Merighi \inst{8}, 
S. Paltani \inst{1},
R. Pell\`o \inst{4}, 
A. Pollo \inst{10}, 
L. Pozzetti \inst{8},
M. Radovich \inst{5}, 
G. Zamorani  \inst{8}, 
E. Zucca  \inst{8}
M. Bondi \inst{2}, 
A. Bongiorno \inst{9}
G. Busarello \inst{5}, 
Y. Mellier \inst{6}, 
P. Merluzzi \inst{5}, 
V. Ripepi \inst{5},
D. Rizzo \inst{4} 
          }

   \offprints{O. Le F\`evre}

   \institute{
1. Laboratoire d'Astrophysique de Marseille, OAMP,
Universit\'e de Provence, UMR 6110, Traverse 
    du Siphon-Les trois Lucs, 13012 Marseille, France\\
              email: olivier.lefevre@oamp.fr
         \and
2. Istituto di Radio-Astronomia - INAF, Bologna, Italy
\and
3. IASF - INAF, Milano, Italy
\and
4. Laboratoire d'Astrophysique - Observatoire Midi-Pyr\'en\'ees, Toulouse, France
\and
5.  Osservatorio Astronomico di Capodimonte - INAF, via Moiariello 16, 80131 Napoli, Italy
\and
6. Institut d'Astrophysique de Paris, UMR 7095, 98 bis Bvd Arago, 75014 Paris, France
\and
7. Observatoire de Paris, LERMA, UMR 8112, 61 Av. de l'Observatoire, 75014 Paris, France
\and
8. Osservatorio Astronomico di Bologna - INAF, via Ranzani 1, 40127 Bologna, Italy
\and
9. Universit\`a di Bologna, Dipartimento di Astronomia - Via Ranzani, 1,
I-40127, Bologna, Italy         
\and
10 Osservatorio Astronomico di Brera - INAF,  Via Brera 28, Milan,
Italy
\and
11 Max Planck Institut fur Astrophysik, 85741, Garching, Germany
\and
12 European Southern Observatory, Karl-Schwarzschild-Strasse 2, D-85748
Garching bei München, Germany
}

   \date{Received September 7, 2004; accepted , February 22, 2005}

   \abstract{
This paper presents the ``First Epoch'' sample from 
the  VIMOS VLT Deep Survey (VVDS). 
The VVDS goals, observations, data reduction with the VIPGI pipeline 
and redshift measurement scheme with KBRED are discussed.
Data have been obtained with the
VIsible Multi Object Spectrograph (VIMOS) on the
ESO-VLT UT3, allowing us to observe $\simeq600$ slits simultaneously
at a spectral resolution $R\simeq230$.
A total of 11564 objects 
have been observed in the VVDS-02h and VVDS-CDFS ``Deep'' fields over
a total area of $0.61$deg$^2$, selected solely on the
basis of apparent magnitude $17.5 \leq I_{AB} \leq 24$. 
The VVDS efficiently covers the redshift range $0 < z \leq 5$.
It is successfully going through the ``redshift desert'' $1.5<z<2.2$,
while the range $2.2 < z < 2.7$ remains of difficult access because
of the VVDS wavelength coverage. A total of 9677 galaxies have a redshift
measurement, 836 objects are stars, 90 objects are AGN,
and a redshift could not be measured for 961 objects. There are
1065 galaxies with a measured redshift  $z\geq1.4$.
When considering only the primary spectroscopic targets,
the survey reaches a redshift measurement
completeness of 78\% overall (93\% including less reliable
flag 1 objects), with a spatial sampling of the population
of galaxies of $\sim25$\% and $\sim30$\%
in the VVDS-02h and VVDS-CDFS respectively. 
The redshift accuracy measured 
from repeated observations with VIMOS and comparison
to other surveys is $\sim$276km/s. 

From this sample we are able
to present for
the first time the redshift distribution of a magnitude-limited
spectroscopic sample down to $I_{AB}=24$. 
The redshift distribution N(z) has a median of
$z=0.62$, $z=0.65$, $z=0.70$, and $z=0.76$, for 
magnitude-limited samples with $I_{AB}\leq22.5$, $23.0$, $23.5$ 
and $24.0$ respectively. A high redshift tail above redshift 2 and
up to redshift 5 becomes
readily apparent for $I_{AB}>23.5$, probing the
bright star-forming population of galaxies. 
This sample provides an unprecedented
dataset to study galaxy evolution over $\sim90$\% of
the life of the universe.

   }

\authorrunning{Le F\`evre, O., Vettolani, G., et al.}
\titlerunning{The VIMOS VLT Deep Survey}
 
   \maketitle
%

\section{Introduction}

Understanding how galaxies and large scale structures
formed and evolved is one of the major goals of modern cosmology.
In order to identify the relative contributions of the various
physical processes at play and the
associated timescales, a comprehensive
picture of the evolutionary properties of galaxies and AGN,
and their distribution in space,
is needed over a large volume and a large cosmic 
time base. 

Samples of high redshift
galaxies known today reach several hundred to a few thousand
objects in the redshift range $0.5-4$, and
statistical analysis suffers from small number statistics, small
explored volumes and different selection biases for each sample,
which prevent detailed analysis.
In contrast, in the local universe, large surveys like the 2dFGRS 
(\cite{colless}) and the Sloan SDSS (\cite{aba})
contain from 250000 up to one million galaxies and reach
a high level of accuracy in measuring the fundamental parameters
of the galaxy and AGN populations. 
In a similar way, we need to
gather large numbers of galaxies at high redshifts to accurately quantify
the normal galaxy population through the  
measurement of e.g. the luminosity function, correlation function and
star formation rate, for different galaxy types and in environments
ranging from low density to dense cluster cores. 

We are currently conducting the VIRMOS VLT
Deep Survey (VVDS), a coherent approach to study
the evolution of galaxies, large scale structures and AGN. The 
observational goals are:
(1) a ``wide'' survey: $\sim100000$ galaxies and AGN observed at low 
spectral resolution
$R\simeq230$, in 16 deg$^2$, to a limiting magnitude $I_{AB}=22.5$ 
and reaching redshifts up to
$\sim1.3$; 
(2) a ``deep'' survey: $\sim35000$ galaxies and AGN observed
at $R\simeq230$ in 2 deg$^2$ and brighter than $I_{AB}=24$,
to map  evolution over $0\leq z \leq 5$, or 90\% of the 
age of the universe; 
(3) an ``ultra-deep'' survey: $\sim1000$ galaxies and AGN brighter
than $I_{AB}=25$, this will probe deep into
the luminosity function (3 magnitudes below M* at z=1).
In addition, we intend to conduct a selected high 
redshift cluster survey following up the clusters
identified in the ``wide'' and ``deep'' surveys,
and a high spectral resolution survey, 
$R\sim2500-5000$, on a sub-sample of 10000
galaxies selected from the ``wide'' and ``deep'' surveys.
The observing strategy that we have devised allows 
us to carry out these
goals in an optimized and efficient approach. 

Guaranteed Time Observations allow completion of
$\sim40$\% of the original survey goals. 
A Large Program is being proposed to the European Southern 
Observatory to complete the VVDS as originally
planned.

We describe here the
survey strategy and the status of the observations in the 
first epoch ``Deep'' survey.
We detail the scientific motivation in Section 2, 
the survey strategy in Section 3, the ``first epoch''
observations in Section 4, the pipeline processing of
the VIMOS data with VIPGI in Section 5, the methods followed
to measure redshifts in Section 6 and the ``first epoch''
VVDS-Deep sample in section 7. After a description of the
galaxy population probed by the VVDS-Deep in Section 8,
we present the redshift distribution of 
magnitude-limited samples as deep as $I_{AB}=24$ in Section 9,
and we conclude in Section 10.

We have used a Concordance Cosmology with $h=0.7$, $\Omega_m=0.3$, and
$\Omega_{\Lambda}=0.7$ throughout this paper.


\section{Survey goals}

\subsection{Science background}

The current theoretical picture of galaxy
formation and evolution and of large scale structure growth in the universe
is well advanced. In contrast, while existing observations already
provide exciting views on the properties 
and distribution of high redshift
objects (\cite{lilly95}, \cite{lefevre95a}; 
\cite{steidel96}, \cite{steidel98}; 
\cite{lilly96}, \cite{madau98};
\cite{cimatti}; \cite{ggds}; \cite{combo17}) 
the samples remain too small, incomplete, affected
by large and sometimes degenerate 
errors from photometric redshifts measurements,
or targeted toward specific populations, 
making it hazardous to relate the evolution
observed in different populations at different epochs, study how evolution
depends upon luminosity, type or local environment of the
populations, and compare results to theoretical predictions.
 
The massive efforts conducted by the 
2dF and Sloan surveys (see e.g. \cite{colless}, 
http://mso.anu.edu.au/2dFGRS/;  and  \cite{aba}, 
http://www.sdss.org/) have been aimed at mapping the distribution and 
establish the properties of more than
one million galaxies in our local environment up to redshifts $\sim0.3$. 
As we get to know better our local universe, deep surveys on volumes
comparable to the 2dFGRS or SDSS  at look back times spanning
a large fraction of the age of the universe are required to
give access to the critical time dimension and trace back the 
history of structure formation on scales from galaxies up
to filaments with $\sim100 h^{-1}$ Mpc. 

This reasoning forms the 
basis of the VIRMOS-VLT Deep Survey (VVDS). It is intended to
probe the universe
at increasingly higher redshifts to establish the evolutionary
sequence of galaxies, AGN, clusters and large scale structure,
and provide a statistically robust dataset to challenge current and future
models, from one single dataset.
Unlike all previous surveys at the proposed depth, the VVDS
is based on a simple and easy to model selection function:
the sample is selected only on the basis of I band magnitude,
as faint as $I_{AB}=24$ to cover the largest possible
redshift range. Although
this produces an obvious bias by selecting increasingly
brighter galaxies when going to higher redshifts, this selection
criterion allows us to perform a complete census of the galaxy
population in a given volume of the universe, above a fixed and
well-defined luminosity.

\subsection{Science motivation}

The  VVDS is designed to address the following goals:

(1) {\bf Formation and evolution of galaxies}: 
The goals are to study the evolution of the main population
of galaxies in the redshift range $0<z<5+$ from a complete census 
of the population of galaxies. 
Several indicators of the evolution status of the galaxy population will 
be computed as a function of redshift and spectral type: counts and colors, 
N(z,type), luminosity and mass function (including vs. spectral type, local galaxy 
density), star formation rate from various indicators, spectrophotometric 
properties, merger rate, etc. These indicators will be 
combined to establish the evolutionary properties of galaxies with 
unprecedented statistical accuracy. The 
multi-wavelength approach (radio at VLA, X-ray with XMM, far-IR with 
Spitzer, and UV with Galex) on the 0226-04 and CDFS deep fields 
will allow us to probe galaxy evolution from the signatures of different
physical processes. HST imaging is on-going on the extended 
VVDS-10h field selected for the HST-COSMOS program.

From the main low resolution survey, it will be possible to select
a well defined sub-sample of high redshift galaxies to be observed at high
spectral resolution to determine the fundamental 
plane of elliptical galaxies at redshifts greater than 1, and up to $z\sim2$
for star forming galaxies.

(2) {\bf Formation and evolution of large scale structures}:
The formation and evolution of large-scale structures (LSS) is yet to be 
explored on scales ~50-100Mpc at redshifts $0.5<z<5+$. The goal is here to
map the cosmic web structure back to unprecedented epochs. From this, 
a variety of statistical indicators 
of galaxy clustering and dynamics can be computed, 
including 3D galaxy density maps,
the 2-point correlation functions 
$\xi(r)$ and $\xi(r_{p},\pi)$ and their projections $w_{p}(r_{p})$, 
power spectrum, counts in 
cells and other higher order statistics. 
The direct comparison of the observed probability distribution function
(PDF) of galaxies with the PDF of dark matter halos predicted
in the framework of cosmological models will be used to yield a measurement
of the evolution of galaxy biasing. The dependence of  
clustering evolution on galaxy type and luminosity
will be investigated. The data will enable comparison of the
3D galaxy distribution with the mass maps produced by weak-shear. 
These indicators will be combined 
to form a coherent picture of LSS evolution from $z\sim4-5$, which will be 
compared to current large N-body simulations / semi-analytical model 
predictions.

(3) {\bf Formation and evolution of AGN}:
While the evolution of the bright AGNs is well constrained by the
large SDSS and 2dF AGN surveys out to large redshifts  (\cite{boyle}),
constraints on fainter AGN are weak.  With no a priori selection of the
survey targets based e.g. on image compactness or color, 
the detection of AGNs along with the
main population of galaxies is feasible (\cite{schade}).
This will allow us to study the 
evolution of the AGN population, in particular the evolution of the luminosity 
function of QSOs out to $z\sim5$, and the evolution of the AGN fraction in the 
galaxy population with redshift. We will also establish the clustering 
properties of AGN. The 0226-04 and CDFS deep fields are 
of particular interest because of the multi-wavelength approach including 
X-ray, radio and far-IR observations.

(4) {\bf Formation and evolution of clusters of galaxies}:
The known sample of high redshift clusters of galaxies above z=0.5 is still 
small, and volume-limited samples are difficult to obtain. In the VVDS 
samples, we expect to identify several dozen clusters (depending
on cosmological models), half of them at a 
redshift above 0.5. We will acquire both large-scale velocity 
information from MOS spectra, and detailed core mapping from IFU spectroscopy. 
This will enable us to describe both the dynamical state of the clusters and 
the spectrophotometric evolution of the galaxy population in clusters. The 
combination of these data with XMM data and weak shear 
analysis will be particularly powerful to establish the mass properties. 
We will aim to derive the evolution of the cluster space density.

These scientific goals have been used to establish
the observational strategy described below.

\section{Survey Strategy}

\subsection{VVDS surveys}

The science goals require a large number of objects over large, deep volumes. 
They are all addressed in a consistent way by the following observational strategy:
(1) {\it Wide survey}: 100000 galaxies and AGN  
observed in the 4 survey fields to 
$I_{AB}=22.5$, $R\simeq230$, the survey will sample $\sim20$\% of the galaxies 
to this magnitude; exposure times are 3000 sec.  
(2) {\it Deep survey}: 35000 galaxies and AGN observed in 
the 0226-04 field and CDFS to $I_{AB}=24$, $R\simeq230$, the survey will 
sample 25\% of the galaxies to this magnitude; exposure times are 16200sec;
(3) 1000 galaxies and AGN observed in the 0226-04 field to 
$I_{AB}=25$, $R\sim200$;
(4) 10000 galaxies in all fields to $I_{AB}=22.5$, R=2500-5000;
(5) 50--100 cluster cores in all fields to $I_{AB}=24$, R=1000.

Efficient observations can be
carried out with $\sim500-600$ galaxies observed 
in one single spectroscopic observation for the ``deep'' survey
and about $350-400$ galaxies in the ``wide'' survey.
The sample is designed
to bring down the statistical noise, to measure e.g. the luminosity
function $\phi*$, M*, or $\alpha$ parameters, or correlation 
length $r_0(z)$ to
better than 10\% at any of 7 time steps. 

\subsection{Multi-wavelength surveys}

In addition to the U, BVRI, JK surveys conducted by the VVDS team 
(\cite{lefevre04}, \cite{mccracken03}, \cite{radovich}, \cite{iovino04},
respectively),
multi-wavelength surveys are carried out by the VVDS team or under
data exchange agreements with other teams. Some of the VVDS fields have
been observed at 1.4 Ghz at the VLA (\cite{bondi03}), with XMM (\cite{pierre03}),
by Galex (\cite{arnouts}, \cite{schiminovich}) and are  being 
observed by Spitzer (\cite{londsdale}).

\subsection{The VIMOS instrument}

The VIsible Multi-Object Spectrograph is installed on
the European Southern Observatory (ESO) Very Large Telescope (VLT),
at the Nasmyth focus of the VLT unit telescope 3 ``Melipal''
(\cite{lefevre03}). 
VIMOS is a 4-channel imaging spectrograph, each channel
(a ``quadrant'') covering $ \sim 7 \times 8 $arcmin$^2$ for
a total field of view (a ``pointing'') of $\sim218$ arcmin$^2$. 
Each channel is a complete spectrograph with the
possibility to insert slit masks $\sim30\times30$cm$^2$ each
at the entrance focal plane, broad band filters or
grisms to produce spectra on a $2048 \times 4096$ pixels$^2$
EEV CCD. 

The pixel scale is 0.205 arcsec/pixel, providing excellent
sampling of the Paranal mean image quality and Nyquist 
sampling for a slit 0.5 arcsecond in width. The spectral
resolution ranges from $\sim200$ to $\sim5000$. 
Because the instrument field at the Nasmyth focus 
of the VLT is large ($\simeq1$m), there is no atmospheric 
dispersion compensator. This requires us to limit observations
to airmasses below 1.7.

In the MOS mode of observations, short ``pre-images''
are taken ahead of the observing run. These are cross-correlated
with the user catalog to match the instrument coordinate
system to the astrometric reference of the user catalog.
Slit masks are prepared using the VMMPS tool, with 
an automated optimization of slit numbers and position
(see \cite{bottini}).

\subsection{VVDS fields}

The 4 main fields of the VVDS have been 
selected at high galactic latitude, and are spread around the 
celestial equator to allow year round survey observations. 
Each field is $2\times2$deg$^2$. 
The imaging is  deep enough to select the VIRMOS targets 
at the survey depth without selection effects
(see Section~\ref{imaging}). We 
have included the Chandra Deep Field South (CDFS, \cite{gia}), which 
is the target of Chandra, HST, XMM and Spitzer multi-wavelength observations,
to complement the work of the GOODS consortium (\cite{goods}).

The fields positions and available optical and near-infrared imaging
data are summarized in Table \ref{fields}.

   \begin{table*}
\begin{center}
      \caption[]{VVDS fields: positions and available optical and near-infrared
photometry}
      \[
        \begin{array}{lllcccccc}
           \hline
            \noalign{\smallskip}
            Field      &  \alpha_{2000} & \delta_{2000} & b  & l  & $Survey mode$ & $Multi-wavelength data$ \\
            \noalign{\smallskip}
            \noalign{\smallskip}
            0226-04 &  02h26m00.0s & -04\deg30\arcmin00\arcsec & -58.0 & -172.0 & Deep & B,V,R,I^a  \\
            VVDS-02h  &            &                           &  & &     & J,K^b \\
                    &            &                           &  & &    & U^c \\
            \noalign{\smallskip}
            \hline
            \noalign{\smallskip}
            1003+01 &  10h03m00.0s & +01\deg30\arcmin00\arcsec & 42.6 & 237.8 & Wide & B,V,R,I^a \\
                    &            &                           &  & &    & J,K^b \\
            \noalign{\smallskip}
            \noalign{\smallskip}
            1400+05 &  14h00m00.0s & +05\deg00\arcmin00\arcsec & 62.5 & 342.4 & Wide & B,V,R,I^a \\
                    &            &                           &  & &    & J,K^b \\
            \noalign{\smallskip}
            \noalign{\smallskip}
            2217+00 &  22h17m50.4s & +00\deg24\arcmin00\arcsec & -44.0 & 63.3 & Wide & B,V,R,I^a \\
                    &            &                           &  & &    & J,K^b \\
            \noalign{\smallskip}
            \noalign{\smallskip}
            CDFS &  03h32m28.0s & -27\deg48\arcmin30\arcsec & -54.5 & 223.5 & Deep & B,V,R,I^d \\
            VVDS-CDFS  &            &                           & & &     & HST B,V,R,I^e \\
            \noalign{\smallskip}
            \hline
         \end{array}
      \]
\begin{list}{}{}
\item[$^{\mathrm{a}}$] (\cite{lefevre04})
\item[$^{\mathrm{b}}$] (\cite{iovino04})
\item[$^{\mathrm{c}}$] (\cite{radovich})
\item[$^{\mathrm{d}}$] (\cite{EIS})
\item[$^{\mathrm{e}}$] (\cite{goods})
\end{list}
\label{fields}
\end{center}
   \end{table*}

\subsection{Field coverage: VIMOS pointing layout for the VVDS-Deep}

%
%
%
%

\begin{figure*}
   \includegraphics[width=17cm]{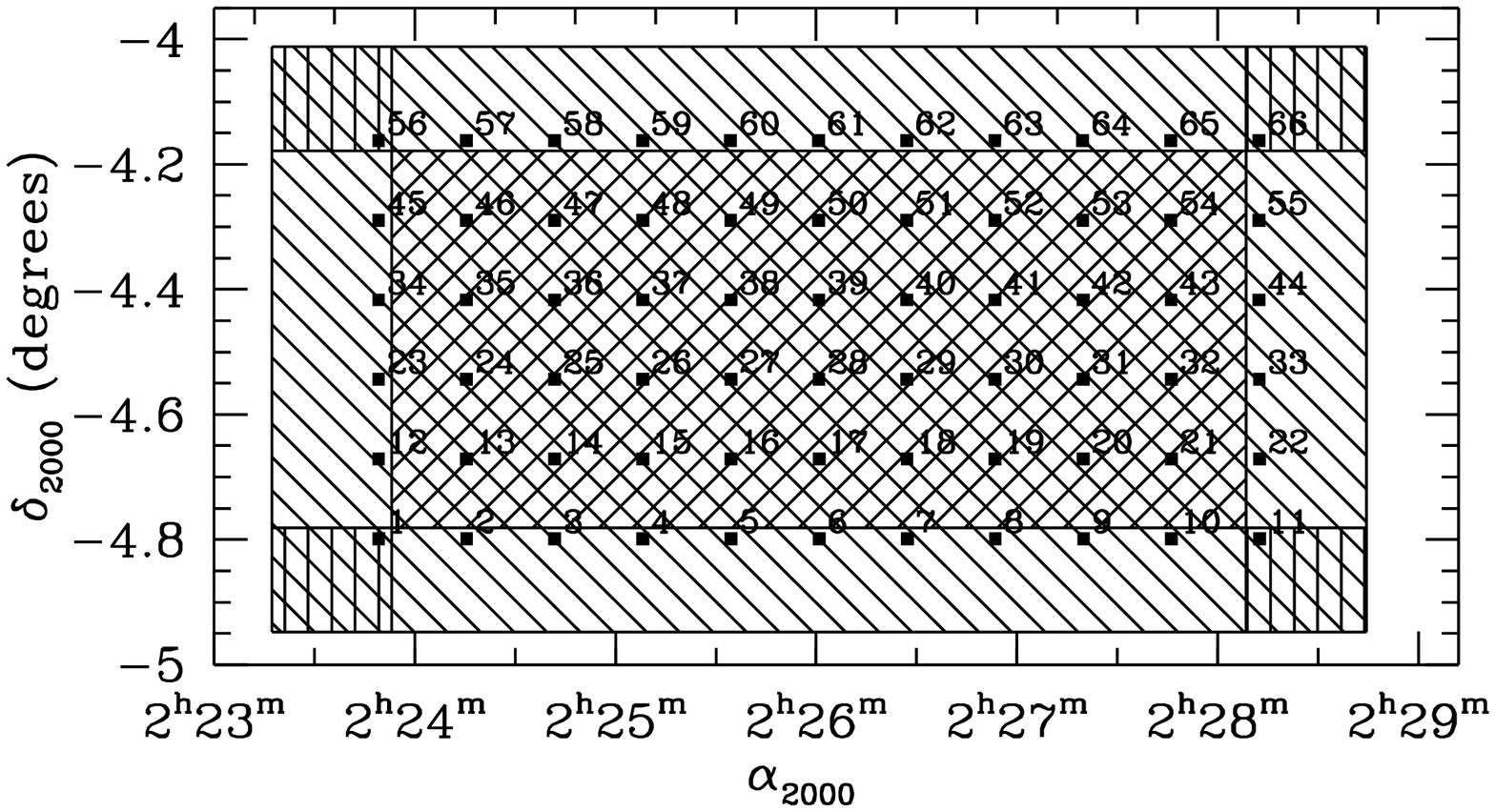}
      \caption{The planned layout and identification
numbers of the VIMOS pointings for the 
VVDS ``Deep'' survey in the VVDS-02h. A total of 66 pointings
is planned, placed on a grid with (2,2) arcmin spacing
in ($\alpha$,$\delta$). This layout allows observation of the
central area (cross hatched) 4 times, and edge areas
twice (hatched) except for the 4 corners which are 
observed once (triple hatched). See Table~\ref{obs} for the list of 
already observed fields and Figure~\ref{xy_02hxy_02h} for the layout of 
objects already observed.
              }
         \label{FigDeep}
   \end{figure*}

The VVDS ``Deep'' survey is conducted in the 
central 1.2deg$^2$ area of the VVDS-02h field and in the 
CDFS. In the VVDS-02h area, we have devised a scheme which allows
uniform coverage of a central area of $0.92 \times 1.33$deg$^2$,
with 4 passes of the instrument, i.e. each point on the
sky has 4 chances of being selected for spectroscopy, using 66
pointings spaced by (2arcmin,2arcmin) in ($\alpha$,$\delta$)
as shown in Figure \ref{FigDeep}. 

The VIMOS multiplex allows measurement of on average $\sim540$
spectra per pointing at the magnitude $I_{AB}\leq24$; this pointing 
strategy therefore enables observation of $\sim35000$ spectra
in the ``deep'' area.

The pointing layout of the VVDS-Wide survey includes
96 pointings to cover $2\times2$deg$^2$ with a grid
of pointings overlapping twice; it will be presented elsewhere
(Garilli et al., in preparation).

\subsection{I-band magnitude-limited sample}
\label{imaging}

The selection function used to identify targets to
be observed spectroscopically has a deep impact
on the final content of the survey. We have taken 
the classical approach of a pure magnitude selection
of the sample, {\it without any other color or
shape criteria}: the VVDS ``Deep'' survey is
limited to $17.5 \leq I_{AB} \leq 24$ and the ``Wide''
survey is limited to $17.5 \leq I_{AB} \leq 22.5$.
The I band selection is made as a compromise 
to select the galaxy population from the
flux emitted by the ``old'' stellar content above
4000\AA ~up to $z\sim1$. At  redshifts above z=1, this
selection criterion implies that galaxies are selected from
the continuum emission below 4000\AA, going increasingly
towards the UV as redshift increases. 

We do not to impose any shape criteria
on the sample selection. This is dictated by the main 
requirement to keep QSOs, compact AGN,
or compact galaxies in the sample .
The ``star-galaxy'' separation from most photometry
codes is known to be only valid for relatively bright
magnitudes, with the locus of stars and galaxies in
a ``magnitude-shape'' diagram overlapping for the faintest
magnitudes. Eliminating star-like objects on the basis of the shape
of the image profile alone would thus eliminate a significant 
part of the AGN and compact galaxy population with little control over
the parent population or redshift domain which is lost. 
We can test this a posteriori from our unbiased spectroscopic
sample; this will be presented elsewhere.

The magnitude selection sets strong requirements on the 
photometric catalog
necessary as input to select the spectroscopic target list. We have conducted
the CFH12k-VVDS imaging survey at CFHT to cover 
the 4 VVDS fields in B, V, R and I bands (\cite{lefevre04}).
The depth of the imaging survey is $\sim1$ magnitude
deeper than the spectroscopic survey limit, which ensures
that there is no a-priori selection against low
surface brightness galaxies (\cite{mccracken03}). The 
multi-color information is used only after the spectroscopic observations,
e.g. to help determine k-corrections (\cite{ilbert04}).
In addition to the core B,V,R, I photometric data,
U band photometry (\cite{radovich}) and J, K' photometry
(\cite{iovino04}) have been obtained for part of the
fields.


\section{First epoch observations: the VVDS-Deep survey in VVDS-0226-04}

\subsection{Preparation of VIMOS observations: VMMPS and database tools}

The preparation of VVDS observations is done from the selected list
of pointings. The VVDS database implemented under Oracle is
used to extract the list of targets allowed in a given
pointing, ensure a secure follow-up of the work on each pointing
and in particular to treat the overlapping observations.
For each pointing a short R-band image is taken 
with VIMOS ahead of the spectroscopic observations. Upon
loading in the database, this ``pre-image'' is automatically
flat-fielded and the detection algorithm of Sextractor 
(\cite{sextractor}) is run to identify the brightest $\sim80$
sources, and a catalog of bright sources is produced
with coordinates in the VIMOS CCD coordinate system.

\begin{figure}[t]
   \centering
   \includegraphics[width=9cm]{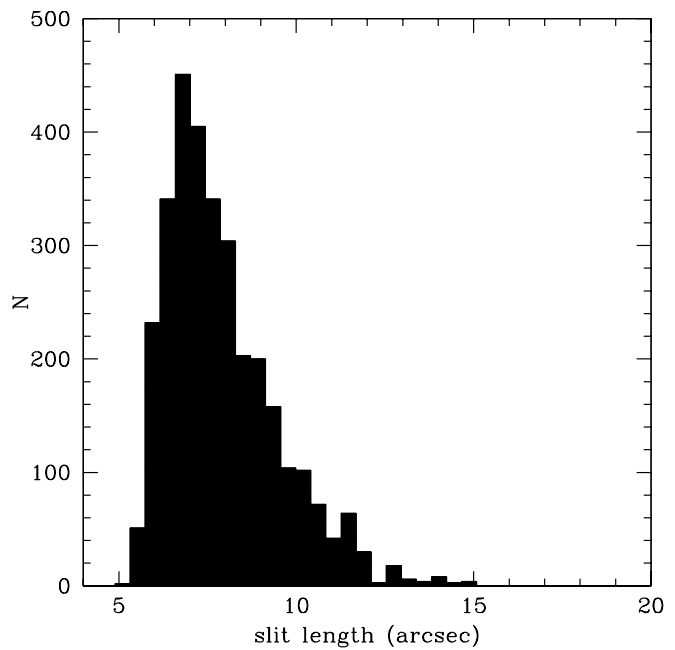}
      \caption{Distribution of slit length in 6 of the
VVDS-02h pointings.
              }
         \label{SlitLength}
   \end{figure}

The next steps are conducted with the VMMPS tool to
design the slit mask layout (\cite{bottini}). 
The positions of the sources identified from the
``pre-image'' are cross-correlated 
with the deep VVDS source catalog (\cite{lefevre04}) to
derive the transformation matrix from the ($\alpha, \delta$) sky
reference frame of the VVDS catalog to the ($X_{CCD}, Y_{CCD}$)
VIMOS instrumental coordinate system. After placing two
reference apertures on bright stars for each pointing
quadrant (see below),
slits are then assigned to a maximum number of 
sources in the photometric catalog. The automated SPOC
(Slit Positioning Optimisation Code, see \cite{bottini}) 
algorithm is run to optimize the number of slits given
the geometrical and optical constraints of the VIMOS
set-up (\cite{bottini}). We have designed masks with slits of one 
arcsecond width, and have forced that a minimum of 1.8 arcseconds
of sky is left on each side of a targeted object to
allow for accurate sky background  fitting and removal 
during later spectroscopic data processing. 
The typical distribution
of slit length in the VVDS is shown in Figure \ref{SlitLength}.
On average, a VVDS-Deep pointing of 4 quadrant masks contains
about 540 slits.

Upon completion of this step, Aperture Definition files
in Pixels (ADP) are produced. ADPs 
contain the positions, length and 
width of all slits to be observed, produced using the VIMOS internal 
transformation matrix from the mask focal plane to
the CCD focal plane to transform the coordinates 
of any VVDS source in the photometric
catalog to VIMOS mask coordinates. The ADPs are then sent to the Mask
Manufacturing Machine (MMU, \cite{conti}) for the masks to
be cut and stored in the VIMOS ``mask cabinets''
ready for observation. Database handling of objects already targeted
ensures  no duplication of observations in overlapping 
areas of adjacent pointings.

\subsection{VIMOS observations}

We have built Observing Blocks (OBs) using the standard ``Jitter'' template,
with 5 steps along the slit, each separated by 0.7 arcsec,
at -1.4, -0.7, 0, 0.7, 1.4 arcsec from the nominal on-target
position. Moving the objects along the slit considerably improves
the data processing in the presence of the strong fringing 
produced by the thin EEV CCDs used in VIMOS (see below).


We have been using the LRRED grism together with the RED
cutoff filter, which limits the bandpass and order overlap. The
useful wavelength range is 5500 to 9400 \AA. With 
one arcsecond slits, the resolution measured at 7500\AA
~is $33$\AA, or $R=227$, and the dispersion is 
$7.14$\AA/pixel. 

\begin{figure*}
\centering
\includegraphics[width=17cm]{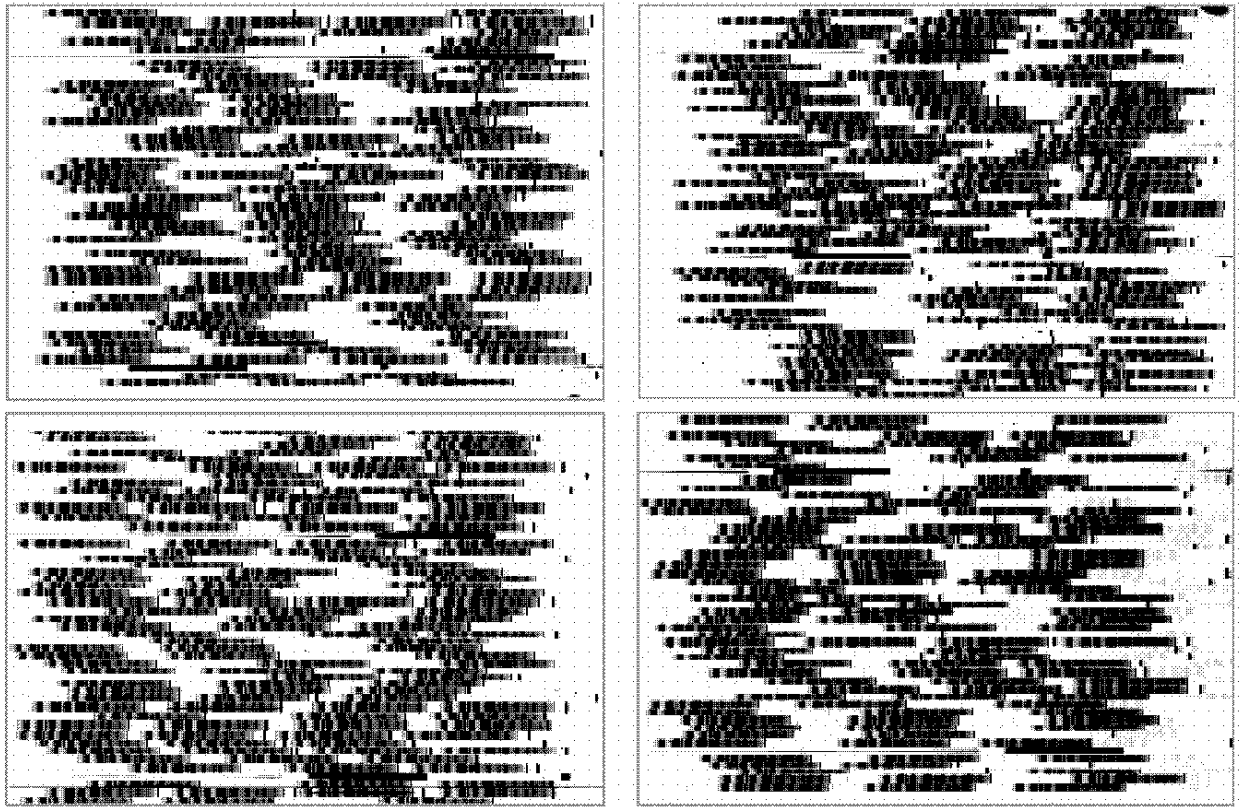}
  \caption{Typical layout of spectra in the 4 quadrants of a VVDS-Wide pointing; more than 550 are observed on average 
          }
\label{specdeep}
\end{figure*}

In the ``Deep'' survey,
10 exposures of 27 minutes each are taken, repeating the
shift pattern twice, for a total exposure time of 4.5h.
The typical layout of spectra on each CCD of the 4 channels is
presented in 
Figure \ref{specdeep} for the ``VVDS-Deep''.


\subsection{Observed fields}

The list of ``VVDS-Deep'' fields observed from October to December 2002 is 
presented in 
Table \ref{obs}. A total of 20 pointings have been observed
in the VVDS-Deep area of the VVDS-02h
field and 5 pointings in the VVDS-CDFS data (\cite{lefevre04}).
We have observed 
$\sim30$\% of the ``Deep'' area pointings during this period.
The first epoch observations of the VVDS-Wide survey will be 
presented elsewhere (Garilli et al., in preparation).

   \begin{table*}
      \caption[]{First epoch VVDS-Deep fields observations in the VVDS-02h and VVDS-CDFS}
      \[
        \begin{array}{lllc}
           \hline
            \noalign{\smallskip}
            $Field$      &  \alpha_{2000} & \delta_{2000} & $Exposure time (min)$  \\
            \noalign{\smallskip}
            \hline
            \noalign{\smallskip}
            F02P016 &  02h25m34.49s & -04\deg40\arcmin15.6\arcsec & 10\times27 \\
            F02P017 &  02h26m00.83s & -04\deg40\arcmin15.6\arcsec& 10\times27 \\
            F02P018 &  02h26m27.16s & -04\deg40\arcmin15.6\arcsec & 10\times27 \\
            F02P019 &  02h26m53.50s & -04\deg40\arcmin15.6\arcsec & 10\times27 \\
            F02P020 &  02h27m19.83s & -04\deg40\arcmin15.6\arcsec & 10\times27 \\
            F02P027 &  02h25m34.47s & -04\deg32\arcmin37.6\arcsec & 13\times27 \\
            F02P028 &  02h26m00.80s & -04\deg32\arcmin37.6\arcsec & 10\times27 \\
            F02P029 &  02h26m27.13s & -04\deg32\arcmin37.6\arcsec & 9\times24 + 5\times27 \\
            F02P030 &  02h26m53.46s & -04\deg32\arcmin37.6\arcsec & 10\times27 \\
            F02P031 &  02h27m19.79s & -04\deg32\arcmin37.6\arcsec & 9\times27, $except channel 2$, 8\times27 \\
            F02P038 &  02h25m34.45s & -04\deg24\arcmin59.6\arcsec & 11\times27, $except channel 3, $ 4\times27 $ (not used)$ \\
            F02P039 &  02h26m00.77s & -04\deg24\arcmin59.6\arcsec & 10\times27 \\
            F02P040 &  02h26m27.10s & -04\deg24\arcmin59.6\arcsec & 11\times27 \\
            F02P041 &  02h26m53.42s & -04\deg24\arcmin59.6\arcsec & 10\times27  \\
            F02P042 &  02h27m19.75s & -04\deg24\arcmin59.6\arcsec & 7\times27, $except channel 2$, 2\times27 $not used$  \\
            F02P043 &  02h27m46.07s & -04\deg24\arcmin59.6\arcsec & 12\times27 \\
            F02P050 &  02h26m00.74s & -04\deg17\arcmin21.6\arcsec & 10\times27 \\
            F02P051 &  02h26m27.06s & -04\deg17\arcmin21.6\arcsec & 10\times27  \\
            F02P052 &  02h26m53.39s & -04\deg17\arcmin21.6\arcsec & 10\times27, $except channel 2: no data$   \\
            F02P053 &  02h27m19.71s & -04\deg17\arcmin21.6\arcsec & 10\times27  \\
            CDFS001  & 03h32m28.0s & -27\degr48\arcmin30\arcsec & 10\times27 \\
            CDFS002  & 03h32m37.04s & -27\degr50\arcmin30\arcsec &  8\times27  \\
            CDFS003  & 03h32m18.95s & -27\degr50\arcmin30\arcsec & 9\times27  \\
            CDFS004  & 03h32m37.04s & -27\degr46\arcmin30\arcsec & 12\times27\\
            CDFS005  & 03h32m18.95s & -27\degr46\arcmin30\arcsec & 10\times27  \\            
\noalign{\smallskip}
            \hline
         \end{array}
      \]
\label{obs}
   \end{table*}

\section{Pipeline processing of VIMOS multi-slit data with VIPGI}

\subsection{VIPGI: the VIMOS Interactive Pipeline Graphical Interface}

The pipeline processing of the VVDS data is performed using the
VIMOS Interactive Pipeline Graphical Interface (VIPGI, 
see \cite{franzetti} for a full description).
This pipeline is built from the same individual C code routines
developed by the VIRMOS consortium and delivered to ESO, but
was made autonomous so as not to be dependent on the  
ESO data environment. 


\subsection{Spectra location}

The location of slits is known from the mask design process,
hence, knowing the grism zero deviation wavelength and 
the dispersion curve, the location of spectra is known a priori on the detectors.
However, small shifts from predicted positions are
possible due to the complete manufacturing and observation
process. From the predicted position, the locations of the 
observed spectra are
identified accurately on the 4 detectors and an extraction window
is defined for each slit / spectrum. 

\subsection{Wavelength calibration}

The calibration in wavelength is secured by the observation
of helium and argon arc lamps through the observed masks.
Wavelength calibration spectra are extracted at the same
location as the object spectra and calibration lines are identified
to derive the pixel to wavelength mapping for each slit.
The wavelength to detector pixel transformation is fit using
a third order polynomial, resulting in a mean deviation 
$\sim1$\AA ~r.m.s. across the wavelength range (\cite{franzetti}).


\subsection{Sky subtraction, fringing correction}

This  critical step in the data processing is handled
at two levels. First, a low order polynomial (second order)
is fit along the slit, representing the sky background
contribution at each wavelength position, and
subtracted from the 2D spectrum. All exposures of a sequence
(5 for the ``Wide'', 10 for the ``Deep'' survey) are then combined 
with a 3-sigma clipping algorithm. As the object is moved to different
positions along the slit following the offset pattern, 
the median of the 2D sky subtracted spectra produces a
frame from which the object is eliminated, but that includes
all residuals not corrected by sky subtraction,
in particular the fringing pattern varying with position
across the slit and wavelength. This sky / fringing residual is
then subtracted from each individual 2D sky subtracted
frame; these are shifted following the offset pattern
to register the object at the same position, and the
individual frames are combined with a median, sigma-clipping
(usually $2\sigma$),
algorithm to produce the final summed,
sky subtracted, fringing corrected, 2D spectrum.

\subsection{1D wavelength and flux calibrated spectra}

The last step done automatically by VIPGI is to extract
a 1D spectrum from the summed 2D spectrum, using an optimum
extraction following the slit profile measured in each slit
(\cite{horne}).
The 1D spectrum is then flux calibrated using the ADU
to absolute flux transformation computed from 
the observations of spectrophotometric standard stars.
We have used the star LTT3218 to perform the flux
calibration.

A final check of the 1D calibrated spectra is performed
and the most discrepant features are removed manually,
cleaning each spectrum of e.g. zero order contamination
or negative non-physical features.

\subsection{Spectrophotometric accuracy}

\begin{figure}
   \centering
   \includegraphics[width=9cm]{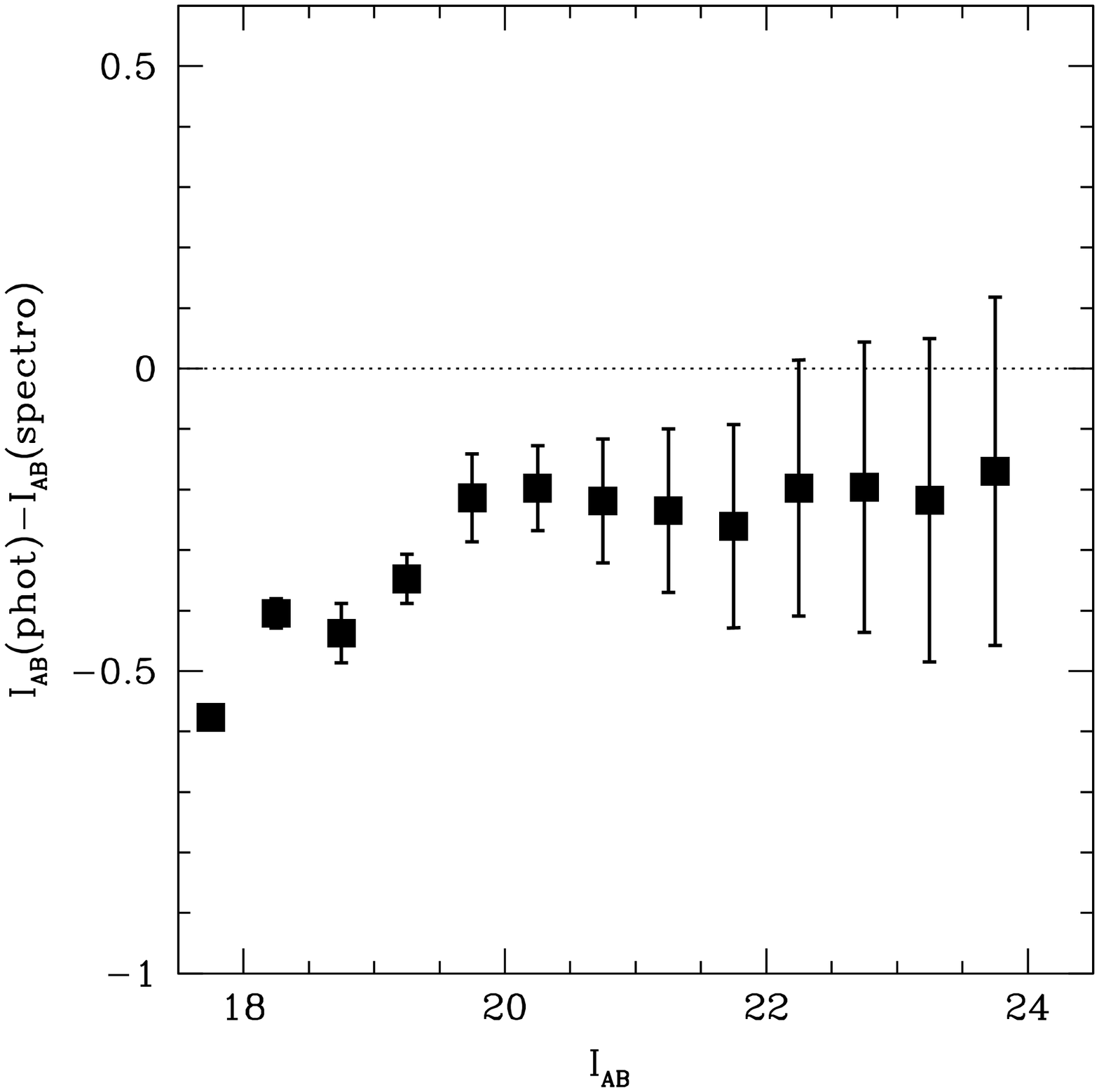}
      \caption{Difference between the photometric I band magnitude
and the I-band magnitude computed from $\sim1000$ calibrated VVDS
spectra. The difference is mainly due to slit losses; no systematic
effect in the spectrophotometric calibration is observed vs. magnitude. 
              }
         \label{phot}
   \end{figure}

The quality of the spectrophotometry calibration can be
estimated by a comparison of integrated magnitudes
computed from the flux calibrated spectra, with the broad band
photometric measurements coming from the deep imaging
data. The comparison shown in Figure \ref{phot}
is done in absolute terms,
comparing the I-band photometric magnitude with the
I-band spectroscopic magnitude derived from the integration
of the flux of spectra in the equivalent of the I-band
photometric bandpass. The spectroscopic I-band magnitude is about 0.2
magnitudes fainter than the photometric magnitude on average,
increasing to about 0.5 magnitudes fainter at the
bright end of the survey. This
can be directly associated with the slit losses inherent to
1 arcsec-wide slits placed on galaxies with sizes larger 
than the slit width. 

\begin{figure*}
   \centering
   \includegraphics[width=15cm]{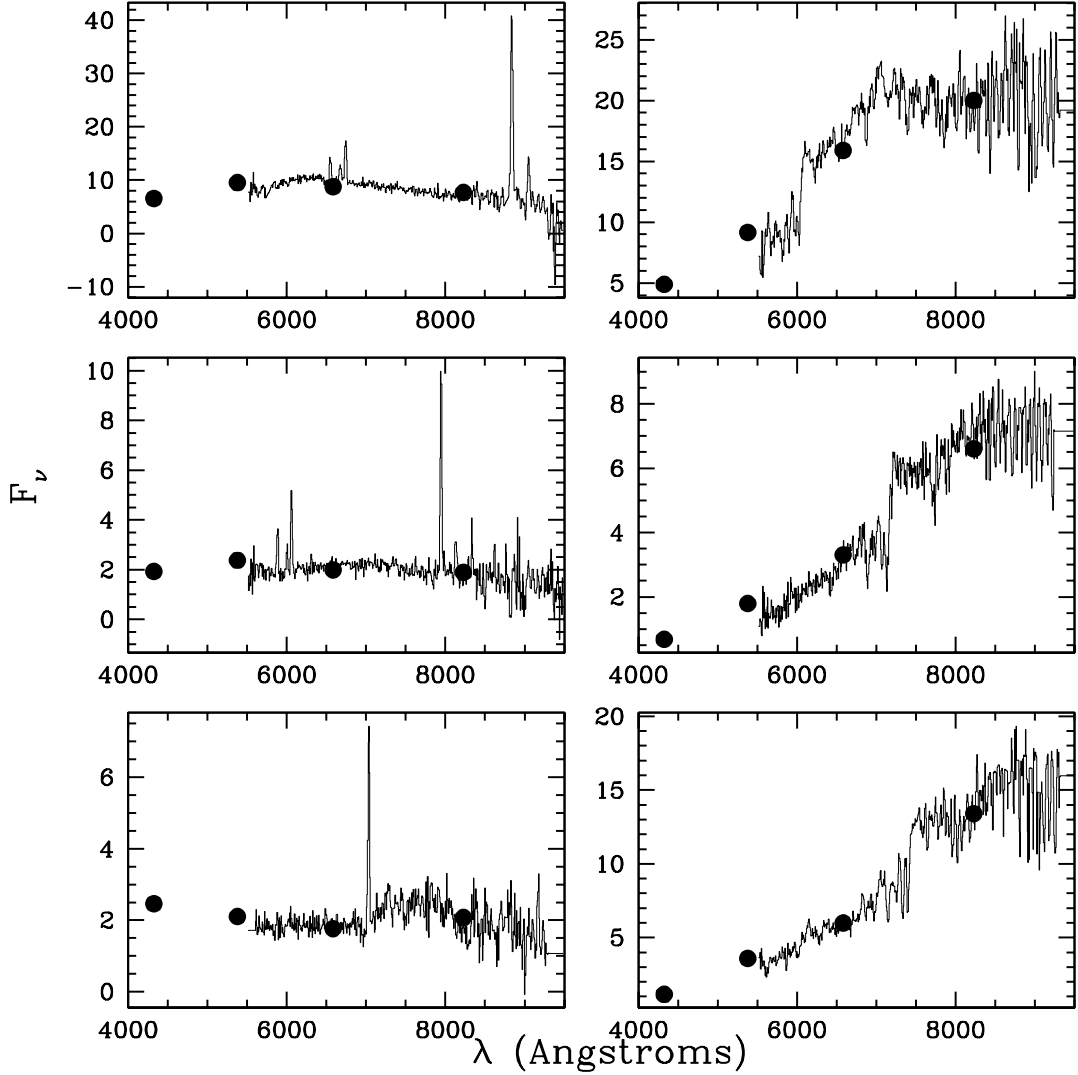}
      \caption{Flux-calibrated VVDS spectra compared to the photometry
of the VVDS imaging survey after scaling to match to 
the I band flux (black dots).
The spectrophotometric accuracy is conserved over the full
wavelength coverage at the 10\% level.             }
         \label{phot2}
   \end{figure*}

The second quality check performed is
the comparison of the photometric multi-color magnitudes
with the flux in spectra, over the full spectroscopic wavelength range.
It is apparent in Figure \ref{phot2} that the relative  spectrophotometry 
is accurate to $\sim10$\% over the
wavelength range $5500-9500$\AA.

\subsection{Spectra signal to noise}

\begin{figure}
   \centering
   \includegraphics[width=9cm]{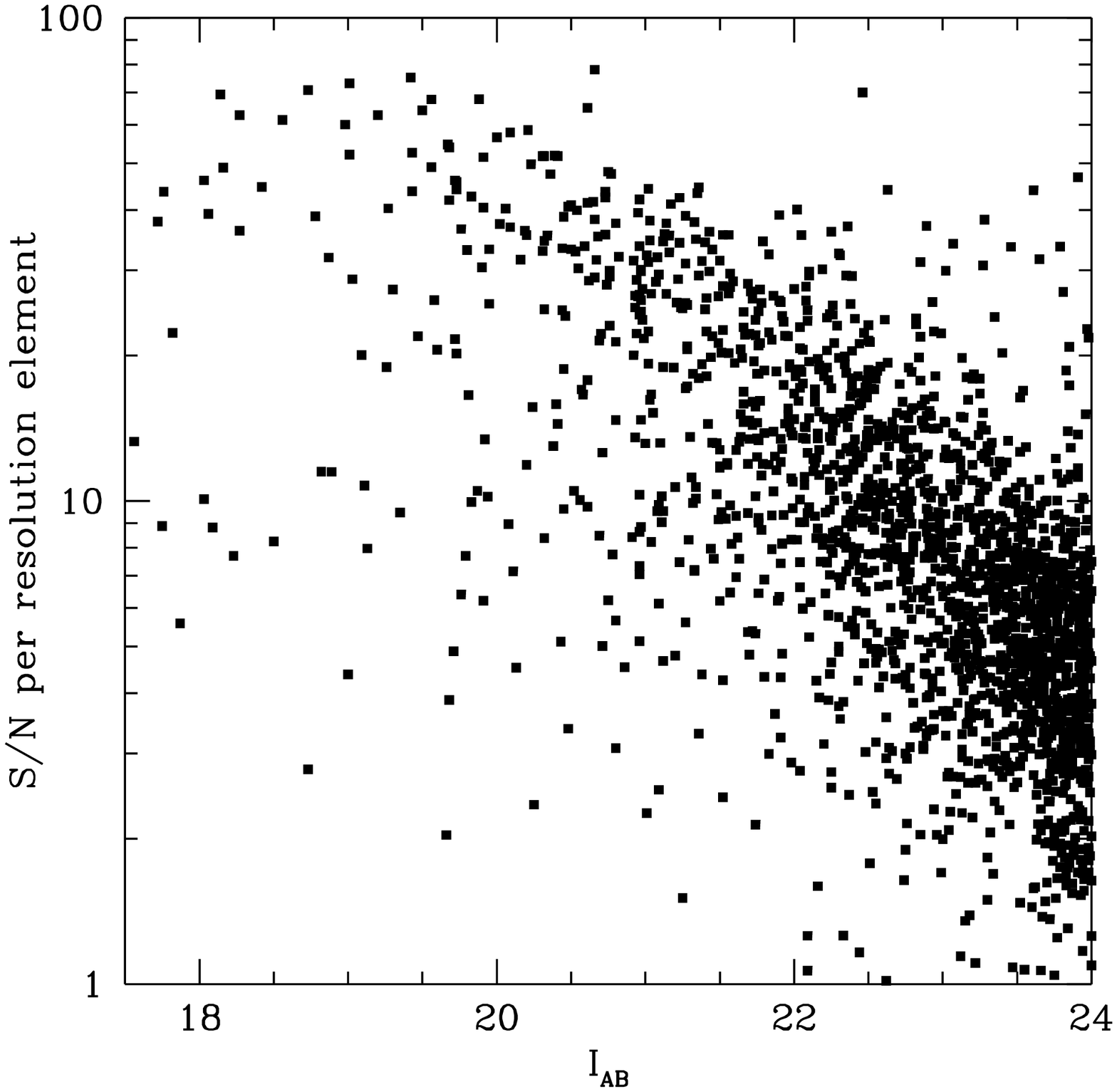}
      \caption{S/N per spectral resolution element over the 
magnitude range of the VVDS-Deep survey.
              }
         \label{sn}
   \end{figure}

The signal to noise ratio, measured per spectral resolution
element of 1000\AA ~centered on 7500\AA, is shown in Figure \ref{sn} for the VVDS-Deep. At the faintest
magnitudes, the S/N goes down to  a mean of $\sim4$.
This S/N is 
sufficient to enable redshift measurements while at the same time
allowing the
observation of a large number of objects as shown
in Sections \ref{s6} and \ref{s7}.

\section{Measuring redshifts}
\label{s6}

\subsection{Building up experience over a large redshift base}

\begin{figure*}
\centering
\includegraphics[width=15cm]{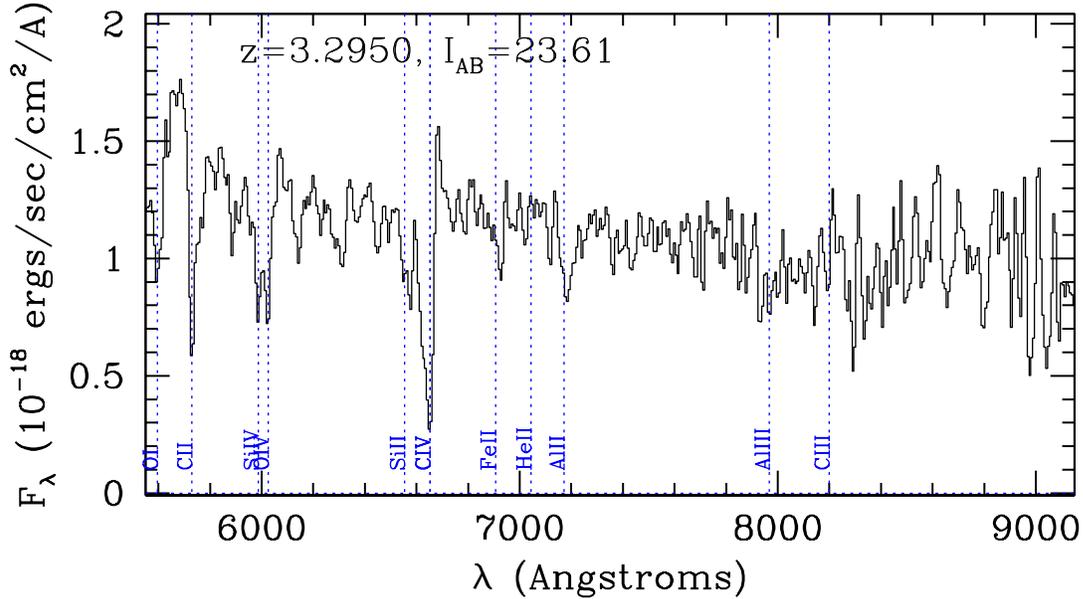}
  \caption{Spectrum of an absorption line galaxy 
with $I_{AB}=23.61$, and $z=3.2950$, demonstrating the
ability of low spectral resolution $R\sim230$
to measure redshifts from absorption line
galaxies down to the very faint end of the survey. 
          }
\label{nice}
\end{figure*}

The VVDS is the first survey to assemble  a 
complete spectroscopic sample of galaxies based on
a simple I-band limit down to $I_{AB}=24$, spanning 
the redshift range $0 \leq z \leq \sim 5$. 
The spectral resolution $R\sim230$ is  adequate
to measure redshifts of absorption line galaxies down to
the faintest magnitudes observed, as shown in Figure \ref{nice}.
Other examples of spectra are given in Section \ref{spec}.

Magnitude-limited samples have
the advantage of a controlled bias in the selection
of the target galaxies, which can lead to a 
secure census of the galaxy population  
as seen at a given rest-frame wavelength (see e.g.
\cite{lilly}). The drawback 
is that, as the magnitudes become fainter, the redshift
range becomes larger, and identifying redshifts
out of a very large range of possibilities becomes
increasingly harder from a fixed set of observed
wavelengths.
The VVDS wavelength range for the VIMOS observations
is $5500-9500\AA$.  It allows a secure 
follow-up of the spectral signature of galaxies 
longward of [OII]3727$\AA$ 
and minimizes the bias in the identification of galaxy 
redshifts up to $\sim1.5$. 

A further difficulty in measuring redshifts in the 
VVDS is that no star/galaxy 
selection is done a priori before spectroscopy, hence
a significant fraction of stars is included in the
target list and they have to be identified in the redshift
measurement process. Redshift
measurements are thus considerably more challenging
than in local surveys or for targeted high redshift surveys
for which the redshift range is known from an a priori
imposed selection function (e.g. Lyman-break galaxies, EROs,
Lyman-$\alpha$ emission objects). 

In addition, measuring the redshifts 
of galaxies with $1.5<z<2.7$ is quite challenging within
the wavelength domain $5500-9500\AA$ that has generally faint
features, and a lack of published observed
galaxy templates in the range $1700-3000\AA$
to be used in cross-correlation programs
such as the KBRED environments developed for the
VVDS (Scaramella et al., in preparation). A dedicated
approach had to be followed to ensure sufficient 
knowledge of rest-frame spectra with VIMOS,
especially in the UV between $\sim1700\AA$ and 
[OII]3727 where previously observed
spectra are not documented in the literature at the S/N level
required to be used as reference templates. These templates
have been combined with
the templates available below $1700\AA$ from the observations of 
Lyman-break galaxies (\cite{shapley}) to produce templates
covering $1100-8000\AA$ rest frame.

\subsection{Crossing the "redshift desert"}

The ``redshift desert'' has been referred to as the
redshift domain between the low--intermediate redshifts
measured up to $z\sim1.3-1.5$ and the high redshift
galaxies identified using the Lyman-break technique,
with a paucity of galaxies identified at $z\sim1.5-2.7$.
Crossing this ``redshift desert'' is critical to reduce the
incompleteness of deep redshift surveys and to probe
the galaxy population at an important time in the
evolution.

In a first pass of redshift measurements on the
first epoch spectra of the 
VVDS $I_{AB}=24$ sample (\cite{lefevre03b}), 
$\sim$10\% of galaxy redshifts appeared hard
to identify while a significant fraction of these spectra
had a good continuum S/N. We have established an iterative
approach to identify these objects based on the 
creation of high S/N rest-frame galaxy templates from observed 
VVDS spectra, going 
as far as possible toward the UV and bridging the gap
between the well observed wavelength
range above [OII]3727$\AA$ and the range below 
1700$\AA$ well represented by the templates assembled 
from $z\sim3$ Lyman-break galaxies (\cite{shapley}). These 
new templates  have been included in 
our redshift measurement engine KBRED
and tested against the parent samples of galaxies used to produce
them, in the range up to redshift $z=1.5$,
then up to redshifts $z=2$ and above up to $z\sim5$ 
as new redshifts were identified.  

This approach has proved successful and has allowed us to
identify a large population of objects with $z>1.5$ as
shown in Sections 8 and 9. Other groups have also recently 
efficiently explored this ``redshift desert'' (\cite{ggds},
\cite{steidel04}). This ``desert'' is now demonstrated
to be only the result of the selection function imposed by the combination
of the faintness of the sources, the wavelength domain of observations,
prior lack of observed galaxy templates and the strong OH sky emission features.
In effect, the ``VVDS redshift desert'' is $2.2 \leq z \leq 2.7$, mainly
associated with the combination of the wavelength domain of spectroscopic
observations and the strong OH sky lines.

A complete account of the templates, the methods to identify the 
galaxies with $z\geq1.5$, and the remaining incompleteness of the VVDS
in the ``redshift desert''  is given elsewhere (Paltani
et al., in preparation).

\subsection{Manual vs. automated redshift measurements with KBRED}

With templates covering the rest-frame range
$1100-8000$\AA, two complementary approaches have been
followed to measure redshifts: a manual approach
and a computer-aided approach with the KBRED automated redshift 
measuring engine (Scaramella et al., in preparation). 

KBRED is a set of routines implemented in the IDL programming 
environment to perform cross-correlation of observed
spectra with reference star and galaxy templates. It includes advanced 
features to perform Principal Component Analysis of the spectra,
projected on a base of reference templates, which are combined to
reproduce the spectral energy distribution and measure the best
redshift. KBRED can be run automatically on a list
of spectra, or manually on a particular spectrum (see 
Scaramella et al. in preparation, for details). 

The first step of redshift measurement has been to run the KBRED 
routine automatically on all 1D spectra, sky corrected and flux calibrated.
This process correctly identifies about 60\% of the redshifts, 
the main difficulty being that, despite intensive testing
and code development, it has not been possible to come up with
a single objective criterion or set of criteria capable of 
identifying a posteriori on which spectra KBRED has 
succeeded or failed to measure
a redshift. This is due to the non-linear noise present in
the spectra, mostly coming from CCD fringing correction residuals 
which can be strong above
$8000\AA$, making the computation of the classical cross-correlation
strength parameter unreliable. Further developments of KBRED
are in progress to solve this difficulty. 

The second step in redshift measurement is then to visually examine 
objects one by one in the VIPGI environment, which makes
available the 1D and 2D corrected spectra, object profile,
sky emission and a set of tools to determine redshifts by hand.
The redshift found by the automatic run of KBRED is displayed
with the main spectral features expected 
superimposed on the 1D spectrum. For $\sim60$\% of the 
spectra the user has simply to quickly verify that the
KBRED redshift indeed corresponds to real spectral features 
and validate the measurement. For the other $\sim40$\%
of the spectra, the user notices that KBRED has identified
a spectral feature that is noise, as determined
upon examination of the 2D spectrum, based on 
the residuals of the sky subtraction at
the location of the strong sky emission OH bands, compounded
by the CCD fringing. A visual examination of the spectrum
is carried out to mark the secure features, removing the
strongest sky-noise features, and either run the VIPGI redshift
calculator matching the marked features, or again
run KBRED on the cleaned spectrum. 

\subsection{Checking the redshift measurements}

We use teams of data reducers to measure 
redshifts. For each pointing set of 4 VIMOS quadrants, 
one person performed the VIPGI processing
from raw data to 2D and 1D sky corrected and calibrated 
spectra. The redshift measurement and quality flag 
assignment was then performed
independently by two persons, and cross-checked
together to solve the discrepant measurements.
The performance of the team in determining the type of spectra, 
the spectral
features to expect and the instrument signatures (e.g. fringing)
increased significantly after the final stage of building
a reference set of
well-defined templates,  
as described above. Therefore, a last
check of the measurements was performed  by a third person, using the
latest set of templates, before they 
were validated and entered in the VVDS database. During this last
``super-check'' the original value was ultimately changed
for about $10$\% of the spectra.

Although time consuming, this procedure ensures that 
minimal machine or personal biases propagate throughout the survey.
The redshift measurements and associated quality flags 
enable a statistical treatment of the overall quality
of the survey, as described below.


\subsection{Quality flags}

We have used a classification of redshift quality similar to
the scheme used in the Canada--France Redshift Survey (\cite{lefevre95a}):\\
-- flag 4: a completely secure redshift, obvious spectral features in
support of the redshift measurement\\
-- flag 3: a very secure redshift, strong spectral features\\
-- flag 2: a secure redshift measurement, several features
in support of measurement\\
-- flag 1: a tentative redshift measurement, weak spectral features
including continuum shape\\
-- flag 0: no redshift measurement, no apparent features\\
-- flag 9: one secure single spectral feature in emission, redshift assigned
to [OII]3727\AA, or $H\alpha$, or in very rare cases to $Ly\alpha$.

A similar classification is used for broad line AGN: when one emission
line is identified as ``broad'' (resolved at the spectral resolution
of the VVDS), flags 11, 12, 13, 14, 19 are used to identify
the redshift quality. At this stage, no attempt has been made
to separate starburst galaxies from type 2, narrow line AGN.

When a second object appears by chance in the slit of the main target,
these objects are classified with a ``2'' added in front of the flag,
leading to flags 20, 21, 22, 23, 24, 29. It is important to identify
these objects separately from the main sample, as a significant fraction
of their flux could be blocked by the slit 
because it has been centered on the main target, hence reducing the
S/N and redshift measurement ability for a given magnitude.

We have classified objects in slits with a clear observational
problem as flag=-10, like e.g. objects for which the automated 
spectra detection algorithm in VIPGI failed, or objects
too close to the edge of a slit to allow for a proper
sky subtraction. This concerns less than 2\% of 
the slits.

\begin{figure*}
\centering
\includegraphics[width=9cm]{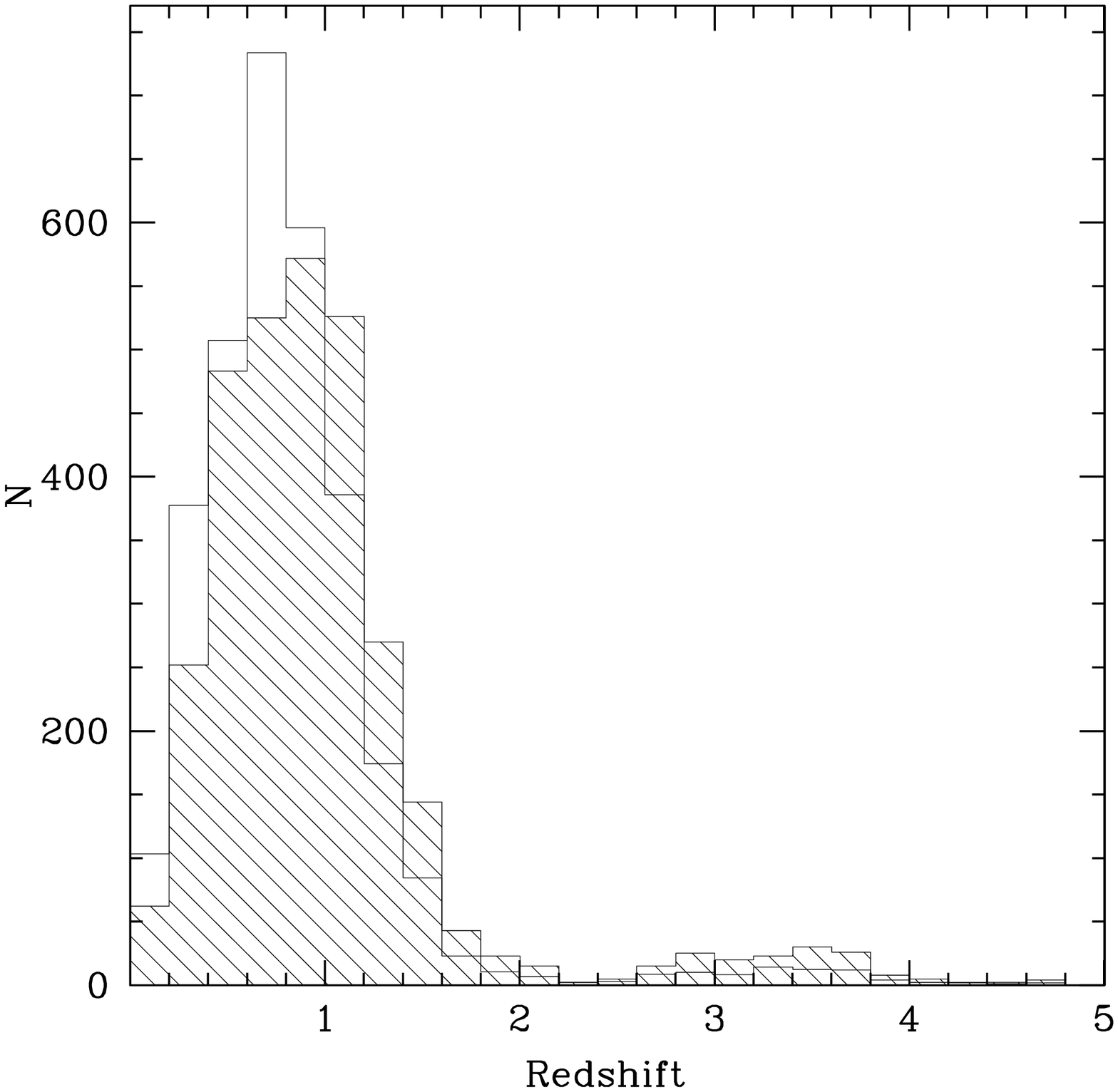}
  \caption{Redshift distribution of flag 2 objects (dashed),
compared to the normalized redshift distribution of the flag 3 and 4
objects (open). 
          }
\label{hist2}
\end{figure*}

\begin{figure*}
\centering
\includegraphics[width=9cm]{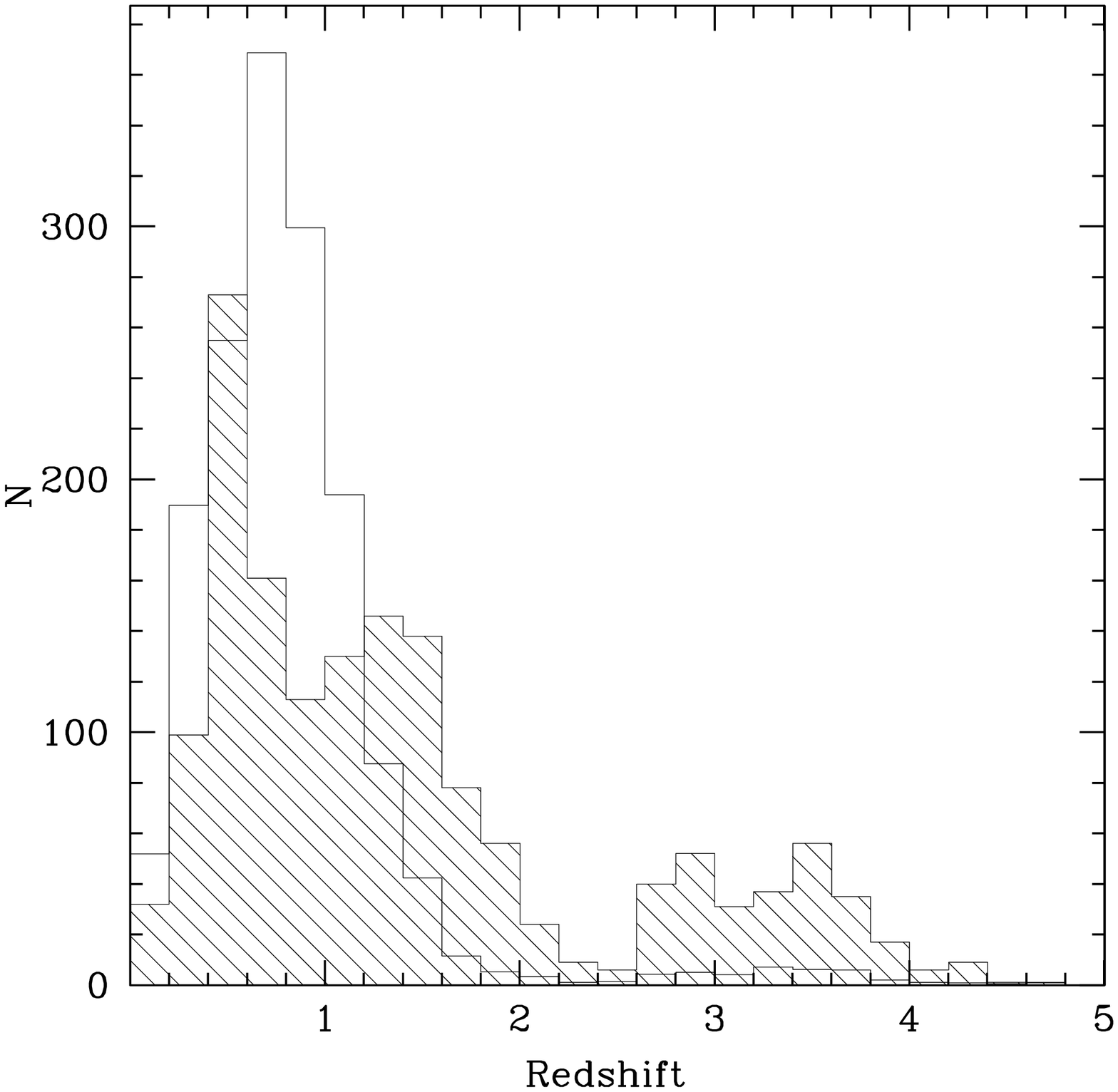}
  \caption{Redshift distribution of flag 1 objects (dashed),
compared to the normalized redshift distribution of the flag 2, 3 and 4
objects (open). 
          }
\label{hist1}
\end{figure*}

The flag number statistics for the VVDS-Deep sample on the CDFS and VVDS-02h
is given in Table \ref{flags}. The redshift distribution of flag 2 
objects reasonably agrees with the overall redshift distribution 
as shown in Figure \ref{hist2}, but only with a 44\% probability
that the two populations are drawn from the same sample (as indicated by
a KS test), with significant differences between the 
two populations for $z>1.2$. The redshift distribution of flag 1 objects is 
significantly different, with a KS test indicating that this population
has only a 7\% probability of being drawn from the same sample as
galaxies with flags 2, 3, and 4.  Flag 1 objects are predominantly 
at $z\geq1.2$ as shown in Figure \ref{hist1}. 

We can estimate the probability of the redshift measurements of being
correct for each of the quality flags in two ways. First we have compared
the difference in redshift for the 426 objects observed twice 
in independent observations: we find a concordance in redshift
within $dz\leq0.0025$,
of $fc=31\pm7$\% for flags 1, $65\pm8$\% for flags 2, $94\pm8$\%
for flags 3, and $99\pm7$\% for flags 4. Assuming that the
intrinsic probability of being correct for a given flag
is a constant $p_{flag}$, then the fraction of concordant 
redshifts with the same flag is $p_{flag}^2$. From the 
fraction of concordant redshifts reported above, we therefore find 
$p_{flag}=fc^{0.5}$, or$p_{flag}=0.55$, $0.81$, $0.97$, $0.995$ for flags 1,
2, 3, and 4, respectively.
 
\begin{figure*}
\centering
\includegraphics[width=15cm]{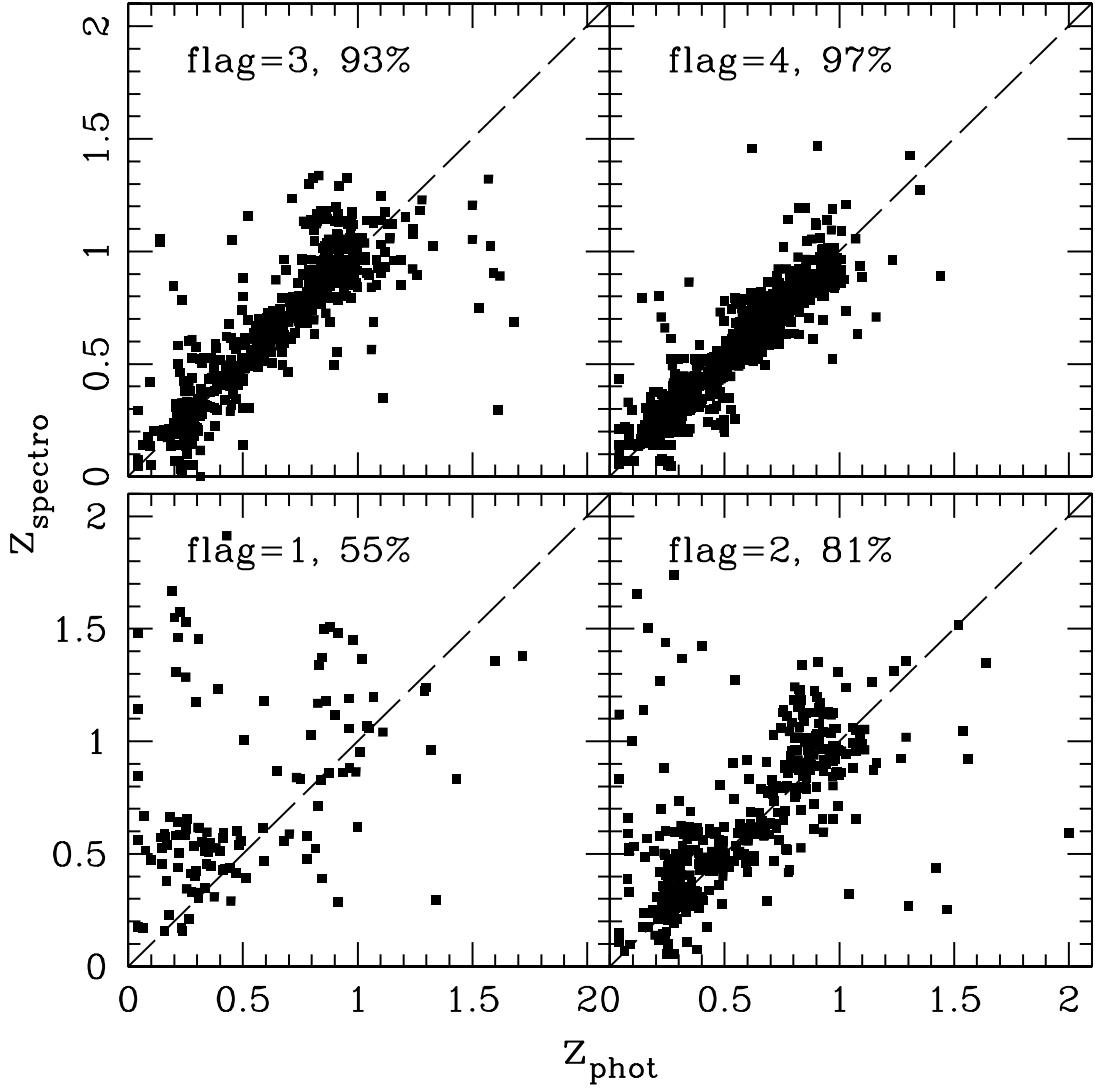}
  \caption{Comparison of spectroscopic redshifts vs. photometric redshifts
computed from BVRI photometry
for objects with magnitude $17.5 \leq I_{AB} \leq 22.5$, for each of 
the quality flags 1, 2, 3 and 4 (from bottom left to top right). The 
fraction of galaxies for which the difference in photometric
redshift vs. spectroscopic redshift is less than 0.2 is indicated
on the top of each panel. 
Flags 3 and 4 represent very secure redshift measurements as
indicated by the comparison of objects observed spectroscopically twice
(see text), hence the dispersion 
observed for these flags is representative of the error in 
measuring photometric redshifts. For flags 2 and 1, errors from the
photometric redshift estimate and the spectroscopic measurements
combine to provide a lower value of 16\% and 42\% of discrepant 
redshifts, respectively. 
          }
\label{zphot1}
\end{figure*}

\begin{figure*}
\centering
\includegraphics[width=15cm]{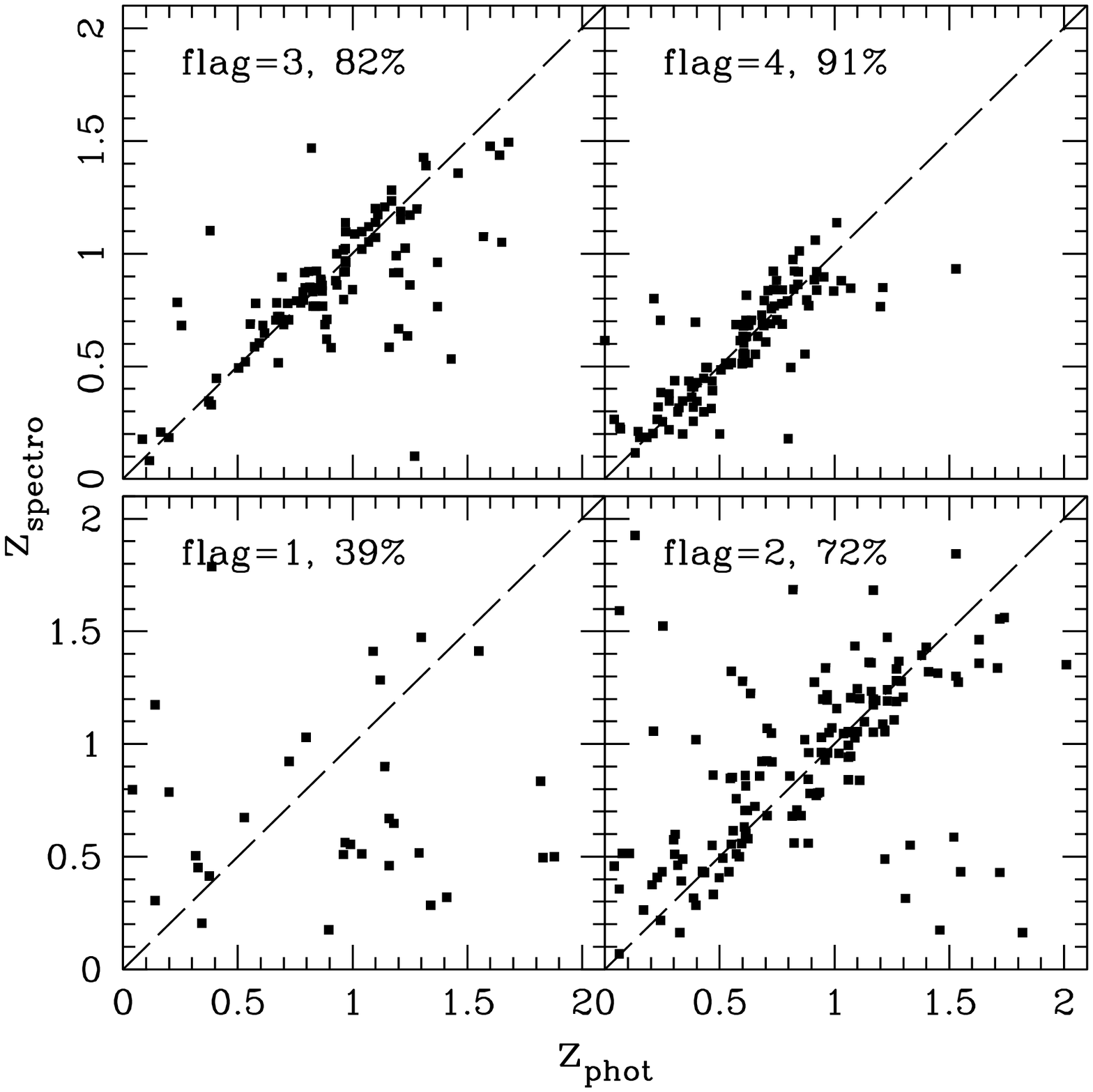}
  \caption{Same as Figure \ref{zphot1} 
for objects with magnitude $22 \leq I_{AB} \leq 24$, for each of 
the quality flags 1, 2, 3 and 4 (from bottom left to top right), 
in the VVDS-02h area where K band data is available in
addition to BVRI data and makes the 
photometric redshifts more accurate. The fraction of concordant
redshifts is 
38\%, 72\%, 82\% and 91\%  for flags 1, 2, 3 and 4 respectively.
Taking into account a photometric redshift failure of $\sim9$\%
as shown by the fraction of concordant flag 4 which are known
to be $\sim100$\% accurate from spectroscopy,
we find that flags 1, 2 and 3 are 47\%, 81\%, 91\% correct, 
respectively,
in this faint magnitude regime where the accuracy 
of photometric redshifts rapidly decreases.
          }
\label{zphot2}
\end{figure*}

A second approach is to
compare the spectroscopic redshifts for the whole sample to 
photometric redshifts derived from the photometric data
(Bolzonela et al., in preparation). To obtain a large number of comparisons for galaxies
with bright magnitudes $17.5 \leq I_{AB} \leq 22$,
we have computed photometric redshifts for the full spectroscopic sample
from the BVRI photometric data. The $z_{phot}$ vs.
$z_{spec}$ comparison in the VVDS-02h field is shown in Figure 
\ref{zphot1}. In addition, to obtain better constraints on photometric redshifts
at faint magnitudes $22 \leq I_{AB} \leq 24$, we have computed 
photometric redshifts for the smaller
sample for which K photometry is available, as shown in
Figure \ref{zphot2}.
From the brightest objects in Figure \ref{zphot1} and assuming that
the flag 4 spectroscopic redshifts are 100\% secure, we can identify an 
intrinsic error of about 5\% in the photometric redshift measurements.
Removing this fraction of failed photometric redshifts, 
we deduce that the spectroscopic flag 3 galaxies
are $\sim96$\% correct, $\sim84$\% for flag 2, while the ``bright''
flag 1 would be $\sim58$\% correct. Using the faintest galaxies
in Figure \ref{zphot2}, we similarly deduce that the 
faint galaxies with flags 3, 2, and
1 are $\sim91$\%, $\sim81$\%, and $\sim48$\% correct, respectively.
We thus determined that
objects with quality flags 1, 2, 3, and 4, are $48-58$\%, $\sim81$\%,
$91-97$\% and $\sim99$\% correct, respectively. This is in
agreement with the values derived from the repeated spectroscopic
observations as reported above.



   \begin{table*}
      \caption[]{Statistics of redshift quality flags for the First Epoch VVDS-Deep sample }
      \[
        \begin{array}{lrrrrrrrrrrrrrrrrrrrrrrr}
           \hline
            \noalign{\smallskip}
$Field$     &  0  &  1   & 2    & 3    & 4    & 9   & 11 & 12 & 13 & 14 & 19 & 20 & 21 & 22 & 23 & 24 & 29 & 211 & 212 & 213 & 214 & 219 & $Total$ \\
            \noalign{\smallskip}
            \hline
            \noalign{\smallskip}
$CDFS$            & 102 &  140 &  506 &  480 &  285 & 115 &  1 &  6 & 1  &  2 &  0 & 20 & 10 & 34 & 10 &  9 & 1  & 0 & 0 & 0 & 0 & 0 & 1722 \\
~~~$galaxies\&QSOs$& -   &  137 &  482 &  458 &  203 & 115 &  1 &  6 & 1  &  2 &  0 &  - & 10 & 31 &  9 &  5 & 1  & 0 & 0 & 0 & 0 & 0 & 1461 \\ 
~~~$stars$        & -   &  3   &  24  &  22  &   82 &   0 &  - &  - & -  &  - &  - &  - &  0 &  3 &  1 &  4 & 0  & - & - & - & - & - & 139 \\ 
$VVDS-02h$        & 690 & 1426 & 2557 & 2187 & 2157 & 304 & 13 & 17 & 27 &  8 & 10 & 149& 73 & 98 & 54 & 41 & 26 & 0 & 2 & 2 & 0 & 1 & 9842 \\ 
~~~$galaxies\&QSOs$& -   & 1363 & 2433 & 2069 & 1776 & 304 & 13 & 17 & 27 &  8 & 10 & -  & 72 & 94 & 52 & 37 & 26 & 0 & 2 & 2 & 0 & 1 & 8306 \\ 
~~~$stars$        & -  & 63   &  124 &  118 &  381 &   0 &  - &  - & -  & -  &  - & -  &  1 &  4 &  2 &  4 & 0  & - & - & - & - & - & 697 \\ 
$Total$     &     &      &      &      &      &     &    &    &    &   &    &     &    &     &    &    &  & & & & & & \\
$VVDS-Deep$ & 792 & 1566 & 3063 & 2667 & 2442 & 419 & 14 & 23 & 28 & 10& 10 & 169 & 83 & 132 & 64 & 50 & 27 & 0 & 2 & 2 & 0 & 1 & 11564 \\
$1st epoch$ &     &      &      &      &      &     &    &    &    &   &    &     &    &     &    &    &   & & & & & & \\ 
            \noalign{\smallskip}
            \hline
         \end{array}
      \]
\label{flags}
   \end{table*}

\section{The ``First epoch'' VVDS-Deep sample}
\label{s7}

\subsection{Galaxies, stars, and QSOs}

A total of 10157 galaxies have a spectroscopically
measured redshift, 8591 for primary targets with flags $\geq2$, and an
additional 1566 with flag 1. An additional 278 galaxies 
with flags $\geq2$ and 83 with flag 1 have been
measured as secondary targets appearing by chance in the
slit of a primary target.

A total of 836 stars have been spectroscopically measured
in the VVDS-02h and the VVDS-CDFS (19 as secondary targets). This 
was expected since no apriori selection was made against
compact objects in the photometric catalog.

There are 71 spectroscopically identified QSOs
(flags 12, 13, 14, 19) ranging in redshift from $z=0.172$
to $z=3.863$, and an additional 14 QSOs with flag 11. 
This unique sample probes the faint
end of the AGN luminosity function at high redshift,
and will be described extensively in subsequent 
papers (Gavignaud et al., Zamorani et al., in preparation).

\subsection{Spatial distribution of observed galaxies}

\begin{figure*}
\centering
\includegraphics[width=15cm]{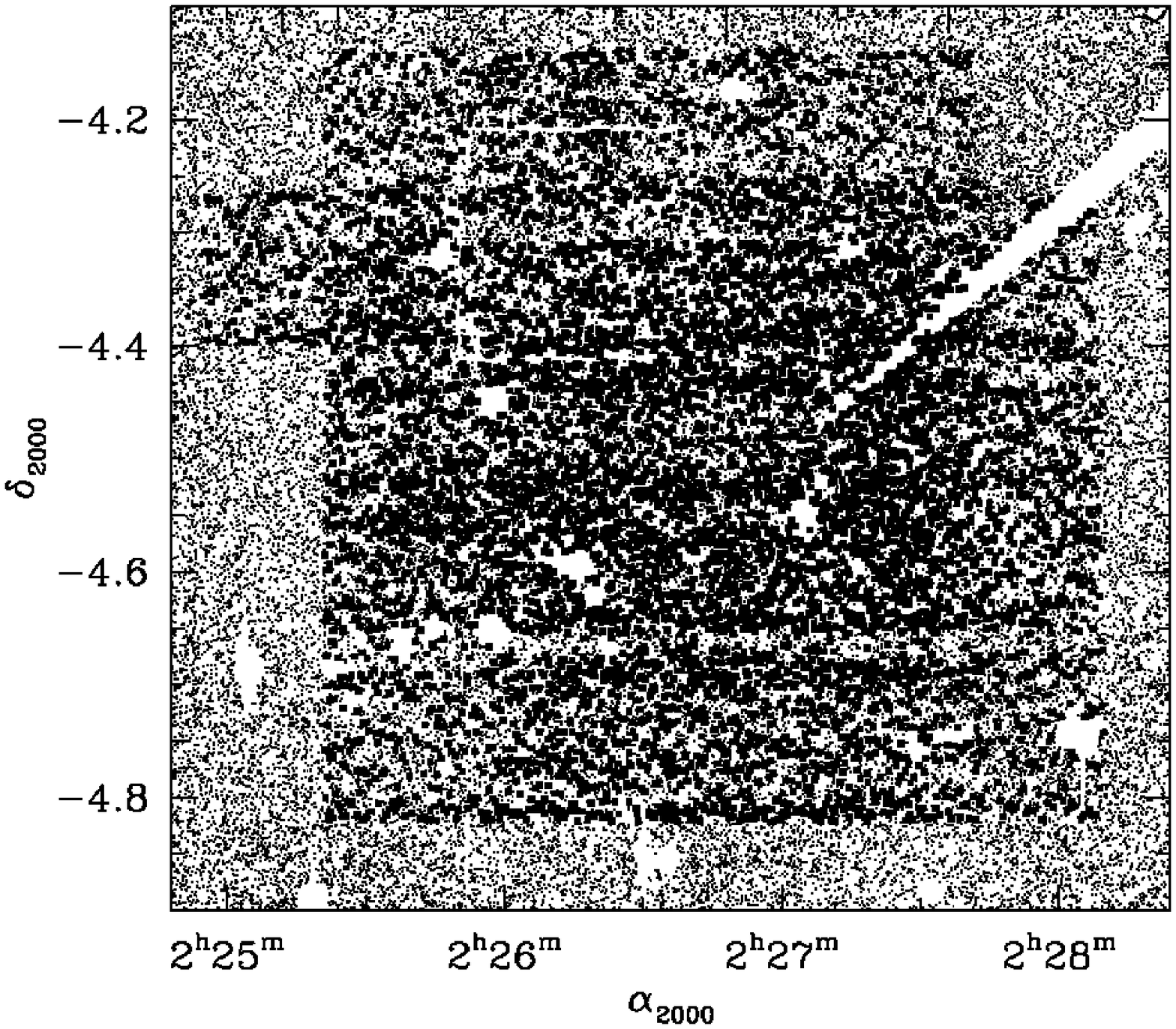}
  \caption{Distribution of galaxies observed in VVDS-02h
          }
\label{xy_02hxy_02h}
\end{figure*}

The spatial distribution of galaxies in the first
epoch observations of the  VVDS-02h field is
shown in Figure \ref{xy_02hxy_02h}, for a total sampled area of
$1750$arcmin$^2$. Together with the VVDS data
obtained in a $453$arcmin$^2$ area around the CDFS (\cite{cdfs}), 
a total $2203$arcmin$^2$, a $0.61$deg$^2$ area has been surveyed. This
constitutes an unprecedented spectroscopic 
survey area at a depth as deep as $I_{AB}=24$.

\subsection{Target sampling rate}

\begin{figure*}
\centering
\includegraphics[width=10cm]{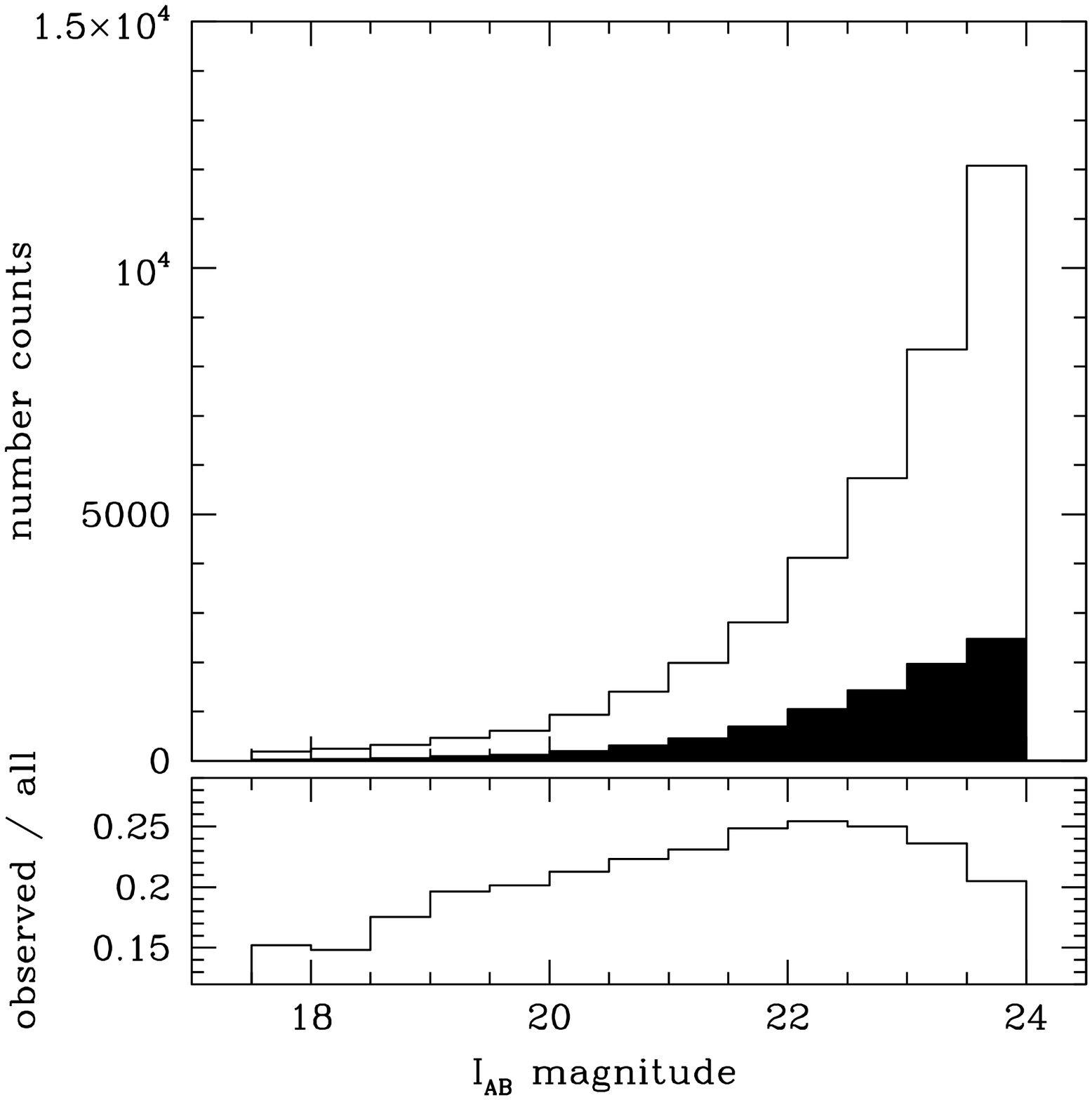}
  \caption{Number and fraction of objects observed in the VVDS-02h,
compared to the total number of objects in the photometric catalog
with $17.5 \leq I_{AB} \leq 24$
          }
\label{distmag}
\end{figure*}

The current sampling of the $17.5 \leq I_{AB} \leq 24$ photometric
sources by the VVDS is indicated in Figure \ref{distmag}
for the VVDS-02h field, while the sampling
in the VVDS-CDFS is almost constant at 30\% (\cite{cdfs}). 
Spectra have been
obtained in the VVDS-02h for a total  
of 22.8\% of the photometric sources, averaged over the
whole area observed, while in the central area corresponding
to about two third of the field, $\sim40$\% of
the photometric sources have been measured. The 
slit optimization technique used in VMMPS favors slit placement
on smaller objects (\cite{bottini}), hence the ratio of spectroscopically 
observed objects to objects in the photometric catalog is
not constant with magnitude for the VVDS-02h, while this
optimization has not been used for the VVDS-CDFS. 

\subsection{Redshift accuracy}

\begin{figure*}
\centering
\includegraphics[width=10cm]{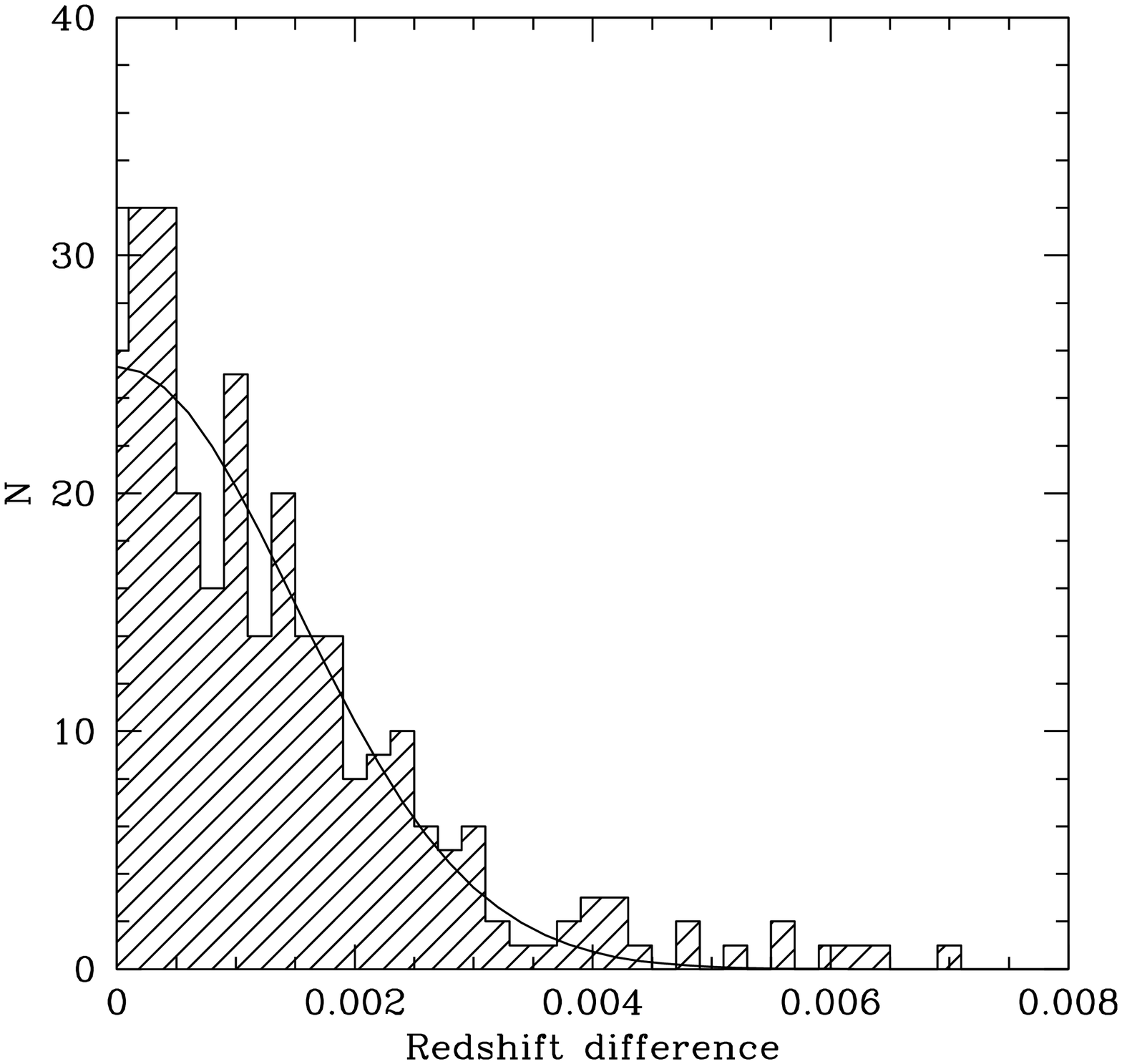}
  \caption{Difference in redshift measurements for 266 galaxies
observed twice in the VVDS-02h and 139 galaxies observed
twice in the VVDS-CDFS. The difference has a 
Gaussian distribution with $\sigma_z=13 \times 10^{-4}$ or 390km/s, 
hence each single
redshift measurement has an accuracy $390/(2)^{0.5}$ or $276$km/s
          }
\label{zacc1}
\end{figure*}

\begin{figure*}
\centering
\includegraphics[width=10cm]{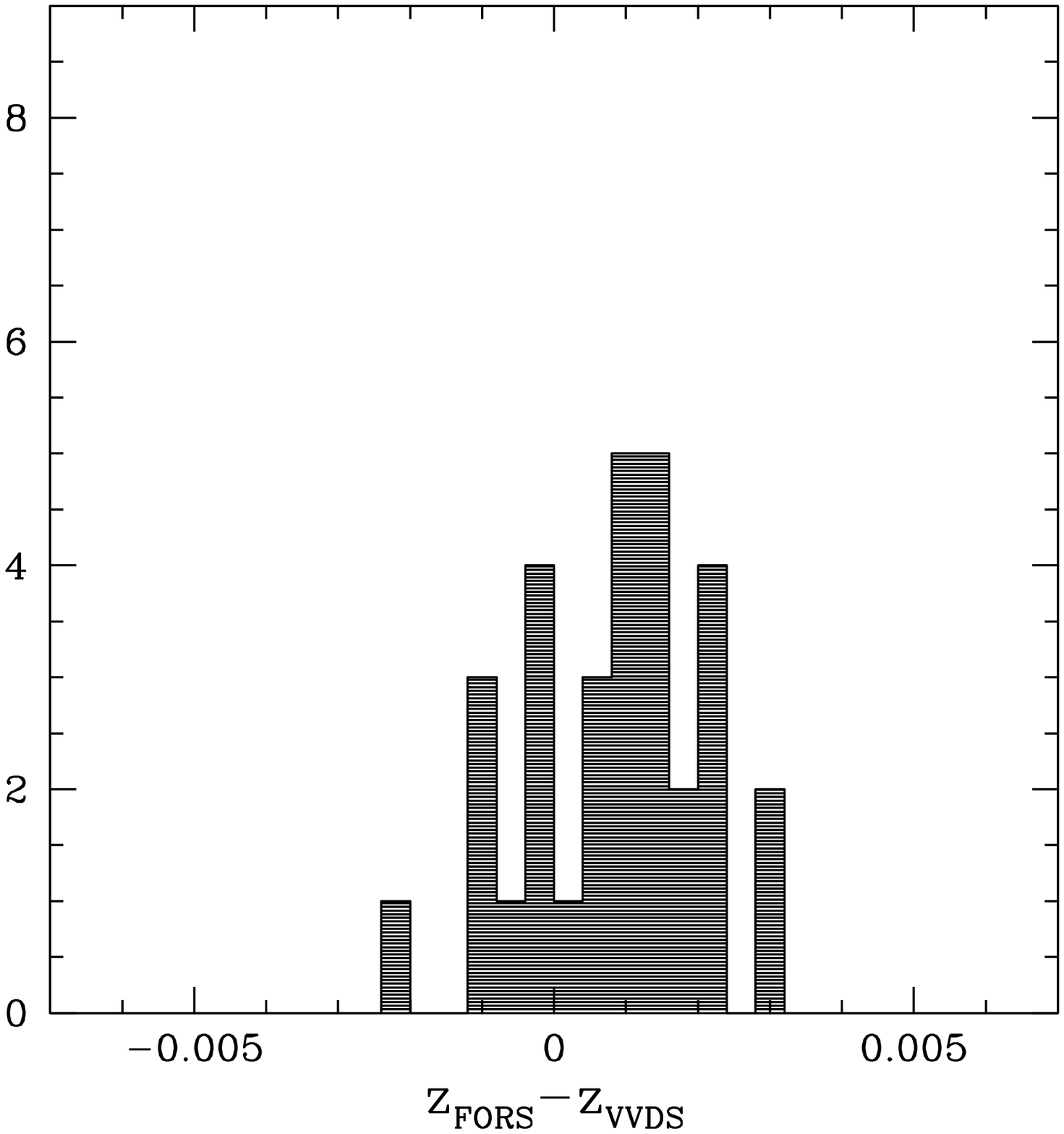}
  \caption{Difference in redshift measurements for 33 objects in
common between the VVDS-CDFS sample and the FORS2-GOODS observations
of \cite{cristiani}. The difference has a mean $dz=8.4.10^{-4}$
and a $\sigma_z=12.6.10^{-4}$ or 378km/s
          }
\label{zacc2}
\end{figure*}

The accuracy of the redshift measurements can be estimated from
the independent observations of the same objects either within
the VVDS or with other instruments. We have observed 160 objects 
twice in the CDFS fields (\cite{cdfs}), and 266 objects twice in
the VVDS-02h. The redshift difference between observations
in the CDFS and in the VVDS-02h
is plotted in Figure \ref{zacc1}; we find that
the difference between two measurements has
a Gaussian distribution with $\sigma_z=13 \times 10^{-4}$,
or $\sim390$km/s, hence the accuracy of single redshift
measurements is $390/(2)^{0.5}$ or $276$km/s. The 33 galaxies observed
in common by VIMOS in the VVDS and FORS2 on the VLT
by the GOODS team in the CDFS (\cite{cristiani}) provide an
external check of our measuring scheme. As shown in Figure \ref{zacc2},
the difference between the two measurements has a
$\sigma_{z_{VVDS}-z_{FORS}}=12.6.10^{-4}$, $378$km/s, very similar to
what is found from the repeat observations in the VVDS.

\subsection{Completeness vs. magnitude}

\begin{figure*}
\centering
\includegraphics[width=12cm]{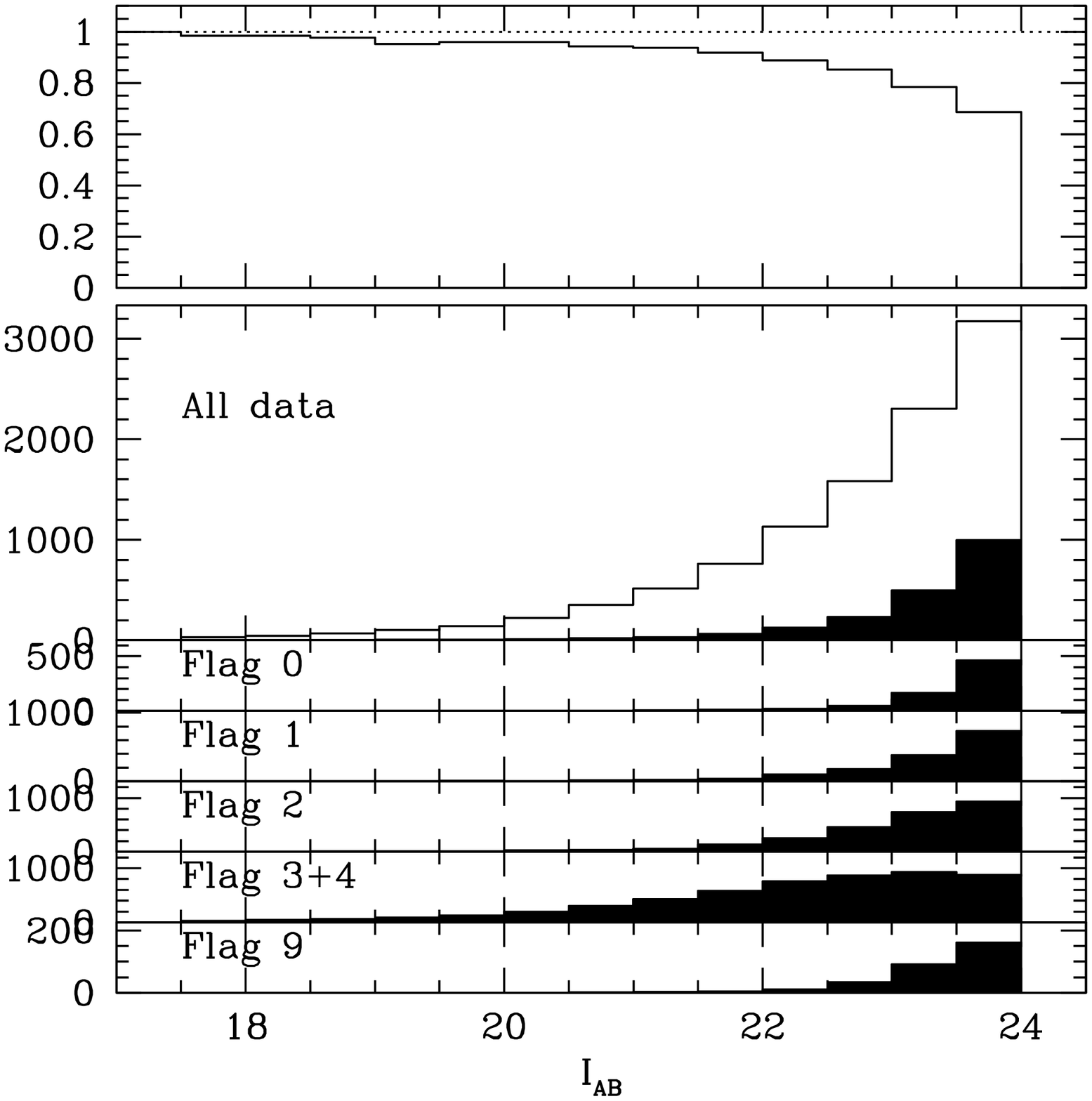}
 \caption{Completeness of the  $I_{AB} \leq 24$ sample in the VVDS-02h
field. (Bottom panel) The magnitude distribution of galaxies
with flags 0, 1, 2, 3+4 and 9;
(Central panel) The magnitude distribution of galaxies with
all flags (open histogram), compared to the distribution
of flags 1 and 2 (filled);
(Top panel) The ratio of secure redshift measurements
with flags 2,3,4,9 and of all measurements (filled). 
The overall redshift measurement
completeness is 78\% (flags 2,3,4,9) and redshifts
are measured for 93\% of the sample (including flags 1).
          }
\label{complete}
\end{figure*}

The completeness in redshift measurement is indicated in Figure \ref{complete}.
Using only the best quality flags (flags 2,3,4,9), the redshift measurement
completeness is 78\%, while including the less secure flag 1 objects 
it reaches 93\%. 
The incomplete fraction translates into a sampling of the galaxy
population which varies with galaxy type and redshift. This is
modeled to compute statistical indicators like
the luminosity function (\cite{ilbert04}).

%
%

\subsection{Spectra}
\label{spec}

\begin{figure*}
   \centering
   \includegraphics[width=\textwidth]{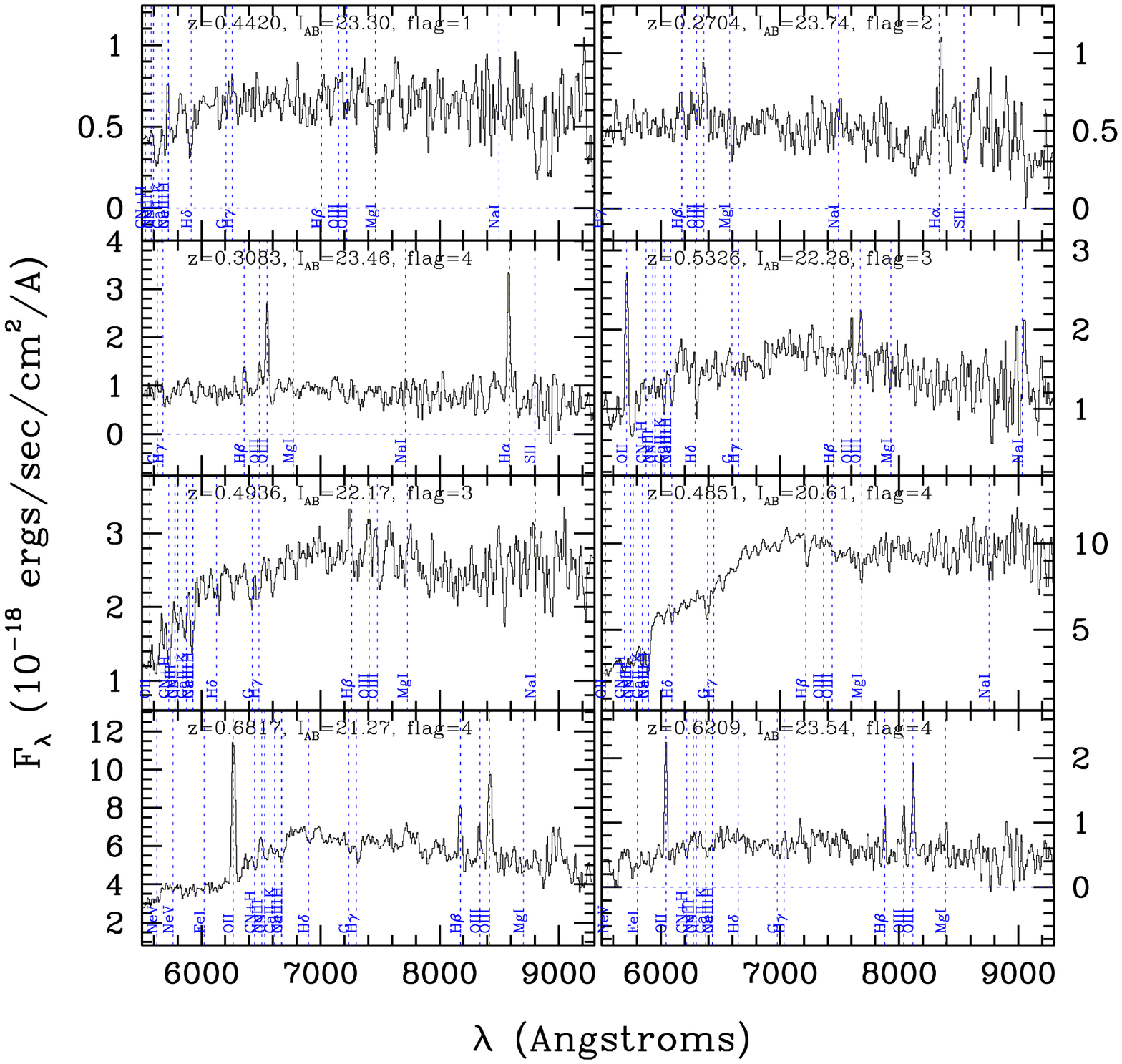}
      \caption{VVDS spectra in the range $0 < z \leq 0.7$ 
              }
         \label{spec1}
   \end{figure*}

\begin{figure*}
   \centering
   \includegraphics[width=\textwidth]{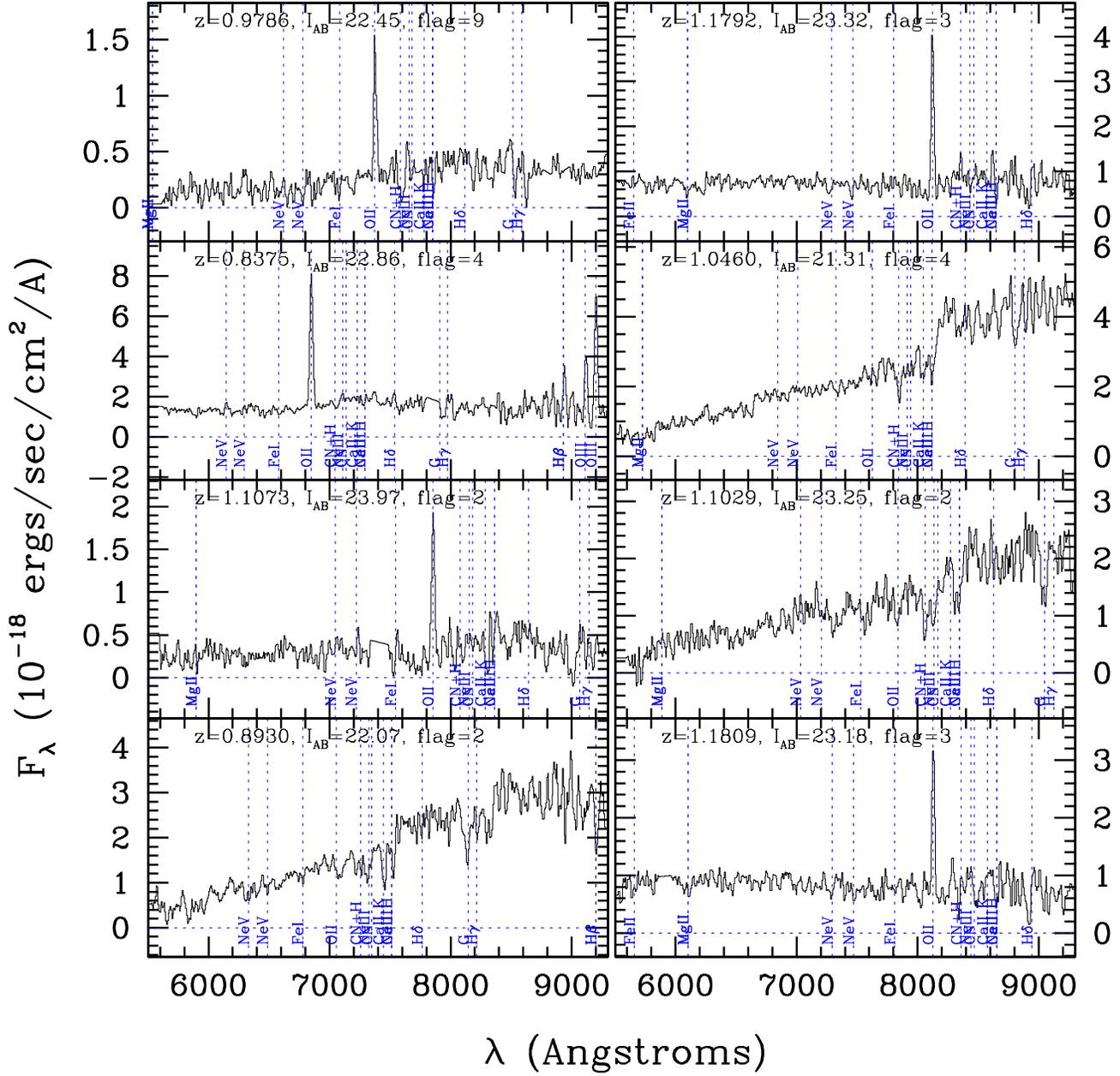}
      \caption{VVDS spectra in the range $0.7 < z \leq 1.3$ 
              }
         \label{spec2}
   \end{figure*}

\begin{figure*}
   \centering
   \includegraphics[width=\textwidth]{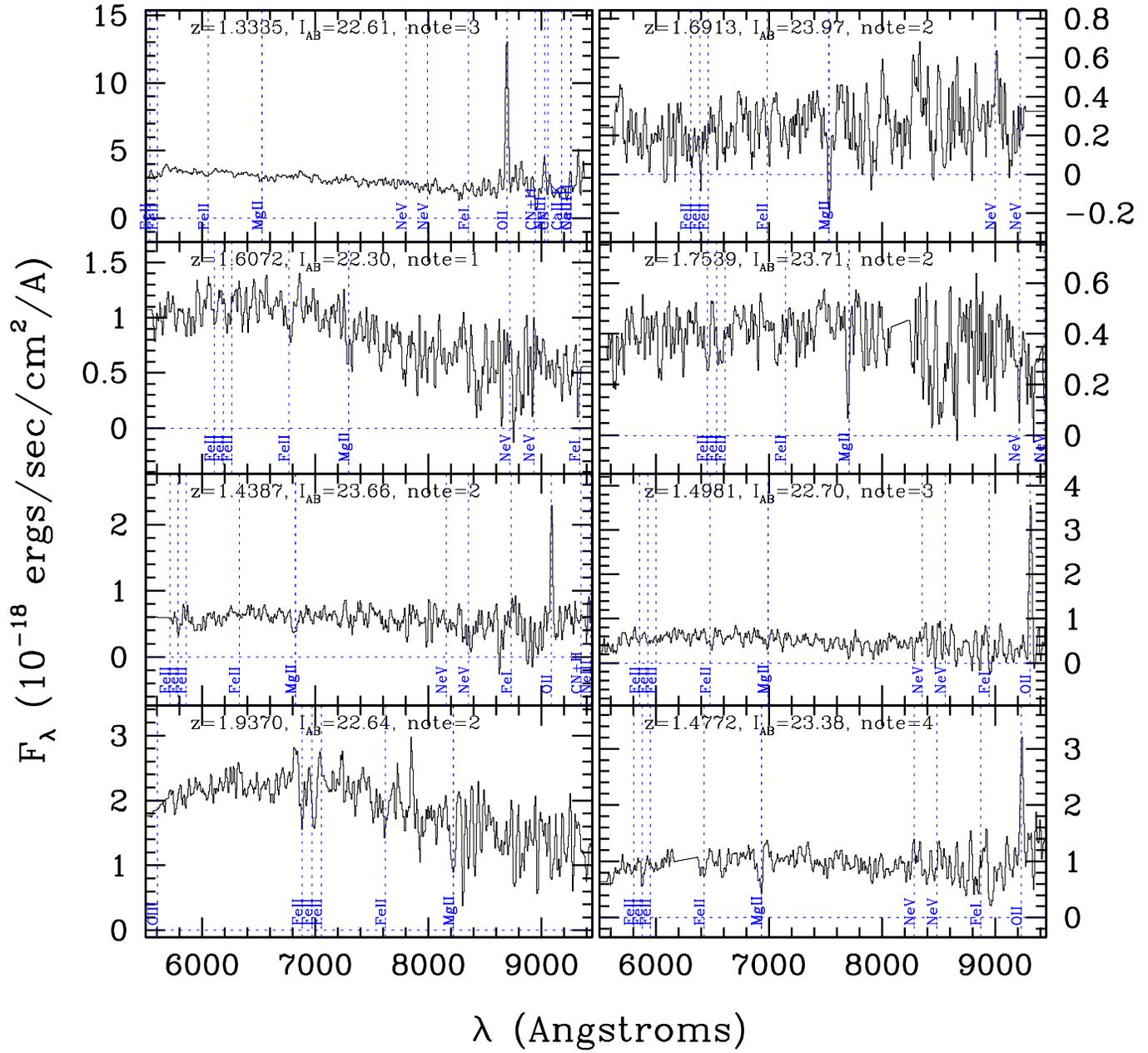}
      \caption{VVDS spectra in the range $1.3 < z \leq 2.2$
              }
         \label{spec3}
   \end{figure*}

\begin{figure*}
   \centering
   \includegraphics[width=\textwidth]{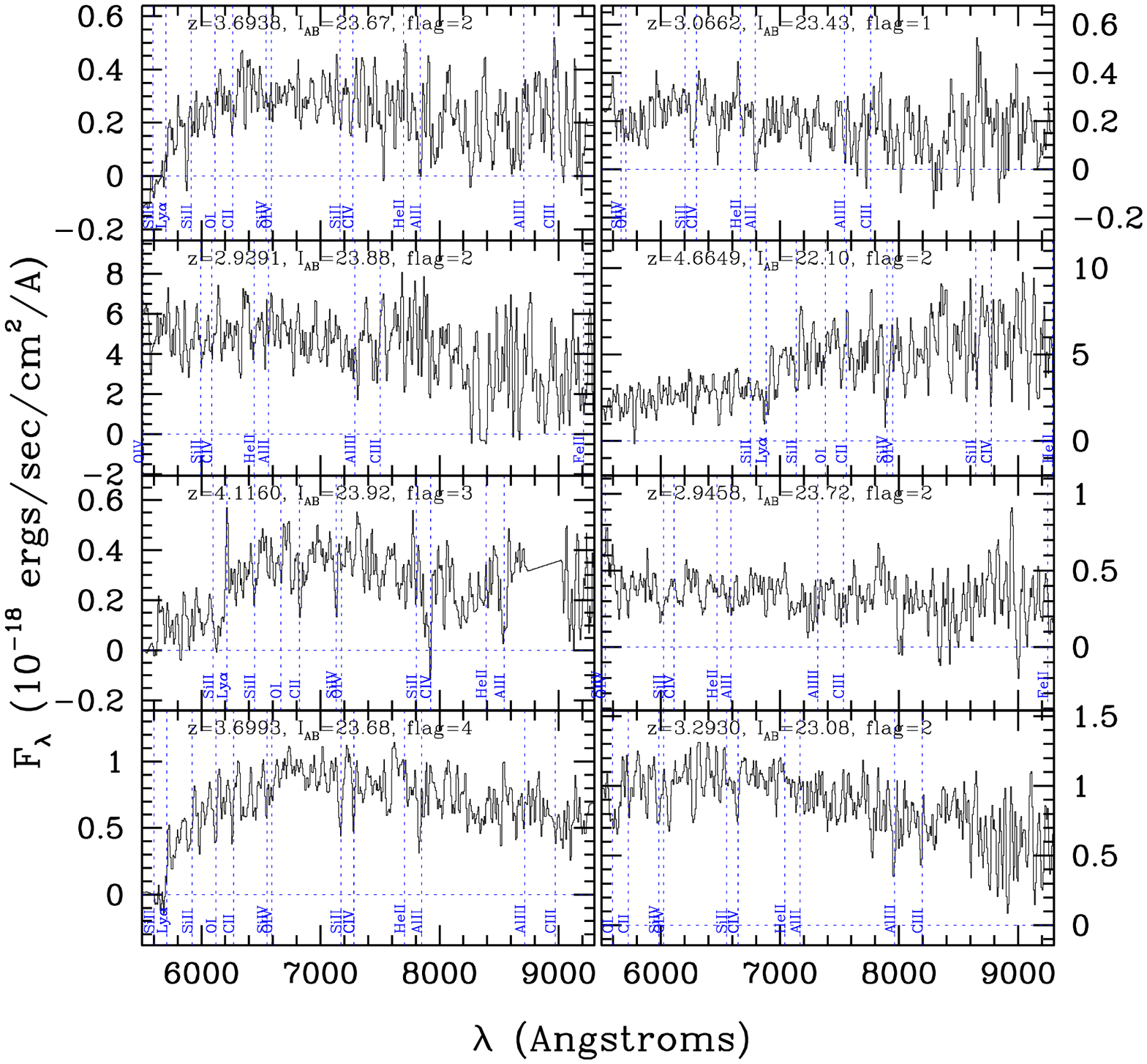}
      \caption{VVDS spectra in the range $2.2 < z \leq 5$ 
              }
         \label{spec4}
   \end{figure*}

Examples of spectra in the various redshift ranges
are given in Figures \ref{spec1} to \ref{spec4}. 

\section{The galaxy population probed by the VVDS-Deep survey}

\subsection{Absolute magnitude distribution}

\begin{figure*}
\centering
\includegraphics[width=9cm]{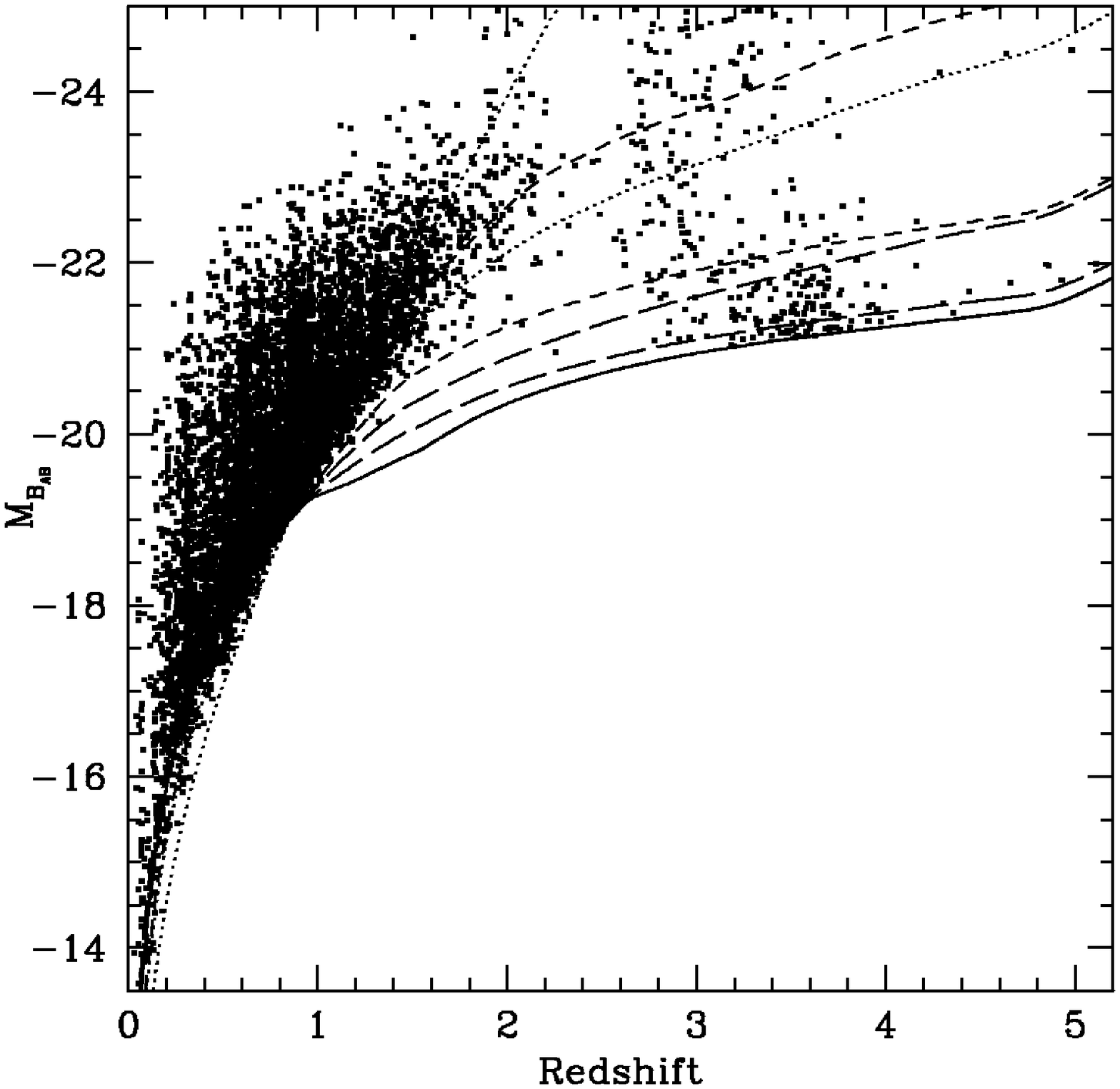}
  \caption{Absolute $M_{B_{AB}}$ magnitude vs. redshift in VVDS-02h.
The full sample is shown together with the tracks of CWW templates
(\cite{coleman}) and starburst templates
used to compute the k(z) correction. At high redshift,
the bluest tracks (starburst galaxies) correspond to the
faintest absolute $M_B$ magnitudes. Each track has
been normalized to produce $I_{AB}=24$
at all redshifts. As redshift
increases, the computation of k(z) for the rest frame B band 
becomes increasingly difficult to constrain, hence produces a
large range in B-band absolute luminosities for $z\geq2$.
At these high redshifts, and given the wavelength
coverage of our spectroscopy and photometry, 
it is more appropriate to compute absolute magnitudes
in the near UV. 
          }
\label{magred}
\end{figure*}

The absolute magnitude $M_{B_{AB}}$  
distribution vs. redshift is presented
in Figure \ref{magred}. At the lowest redshifts $z<0.4$, the VVDS-Deep
observes intrinsically faint galaxies with a mean 
$M_{B_{AB}}\sim-18$ spanning
a large range $-13.5 \leq M_{B_{AB}} \leq -21$. At intermediate redshifts
$0.4<z<0.8$, galaxies are observed with a mean
$M_{B_{AB}}\sim-19.5$ in the range 
$-17 \leq M_{B_{AB}} \leq -22.5$, while for $1<z<2$ observed galaxies are
brighter than $M_{B_{AB}}\sim-19$. This is a consequence
of the pure apparent I-band magnitude selection, and the shift
in the mean and range in absolute luminosity with redshift
has to be taken into account when interpreting the observations.
As the redshift range is large, the k(z) corrections applied to 
transform apparent magnitudes to the rest frame B absolute
magnitude $M_{B_{AB}}$ work best for redshifts $z<1.5$
for which our broad band photometry can be used to constrain
the rest frame B luminosity. At $z>1.5$ the computation
of k(z) and $M_{B_{AB}}$ using template fitting becomes more uncertain, and 
UV rest absolute magnitudes are more appropriate (see
\cite{ilbert04} for more details).


\subsection{Population at increasing redshifts to  $z \sim 5$}

The number of galaxies in several redshift slices is given in
Table \ref{zcounts}. 
There are:\\
-- 4004 objects with $z \leq 0.75$ and secure redshifts (4528
have measured redshifts when flag 1 objects are included)\\
-- 3344 objects with $0.75 < z \leq 1.4$ and secure redshifts (3744
have measured redshifts when flag 1 objects are included)\\
-- 305 objects with $1.4 < z \leq 2.5$ and secure redshifts (603
have measured redshifts when flag 1 objects are included)\\
-- 186 objects with $2.5 < z \leq 5.0$ and secure redshifts (462
have measured redshifts when flag 1 objects are included)

This constitutes  the largest spectroscopic redshift sample
over the redshift range $0.4 \leq z \leq 5$ assembled to
date. A detailed 
account of the properties of the various populations
probed by the VVDS-Deep will be given elsewhere (Paltani
et al., in preparation).

   \begin{table*}
      \centering
      \caption[]{Number of measured galaxies in redshift slices of the VVDS-Deep First Epoch (primary targets only)}
      \[
        \begin{array}{lcccccccc}
           \hline
            \noalign{\smallskip}
$Field$ &  $0-0.5$  & $0.5-0.75$  & $0.75-1$  & $1-1.4$  & $1.4-2.5$ & $2.5-3.5$ & $3.5-5$ & $All z$ \\
            \noalign{\smallskip}
            \hline
            \noalign{\smallskip}
$CDFS, flag$ \geq 1      &  280 &  518 &  245 &  265 &  50 &  22 &  15 & 1395 \\ 
$VVDS-02h, flag$ \geq 1  & 1674 & 2056 & 1801 & 1433 & 553 & 264 & 161 & 7945 \\ \hline
$Total$                  &      &      &      &      &     &     &     & \\
$VVDS-Deep$              & 1954 & 2574 & 2046 & 1698 & 603 & 286 & 176 & 9340 \\
$1st epoch, flag$ \geq 1 &      &      &      &      &     &     &     & \\ \hline
$CDFS, flag$ \geq 2      &  259 &  490 &  223 &  243 &  28 &   8 &  7  & 1258 \\ 
$VVDS-02h, flag$ \geq 2  & 1466 & 1789 & 1688 & 1190 & 277 & 101 & 70  & 6582 \\ \hline
$Total$                  &      &      &      &      &     &     &     & \\
$VVDS-Deep$              & 1725 & 2279 & 1911 & 1433 & 305 & 109 & 77  & 7840 \\
$1st epoch, flag$ \geq 2 &      &      &      &      &     &     &     & \\ \hline
            \noalign{\smallskip}
            \hline
         \end{array}
      \]
\label{zcounts}
   \end{table*}

\section{Redshift distribution for magnitude-limited
samples at $I_{AB}\leq22.5$ and $I_{AB}\leq24$}

\subsection{Redshift distribution for $17.5 \leq I_{AB}\leq22.5$,
$17.5 \leq I_{AB}\leq23$, and $17.5 \leq I_{AB}\leq23.5$ magnitude limited samples}

\begin{figure*}
   \centering
   \includegraphics[width=\textwidth]{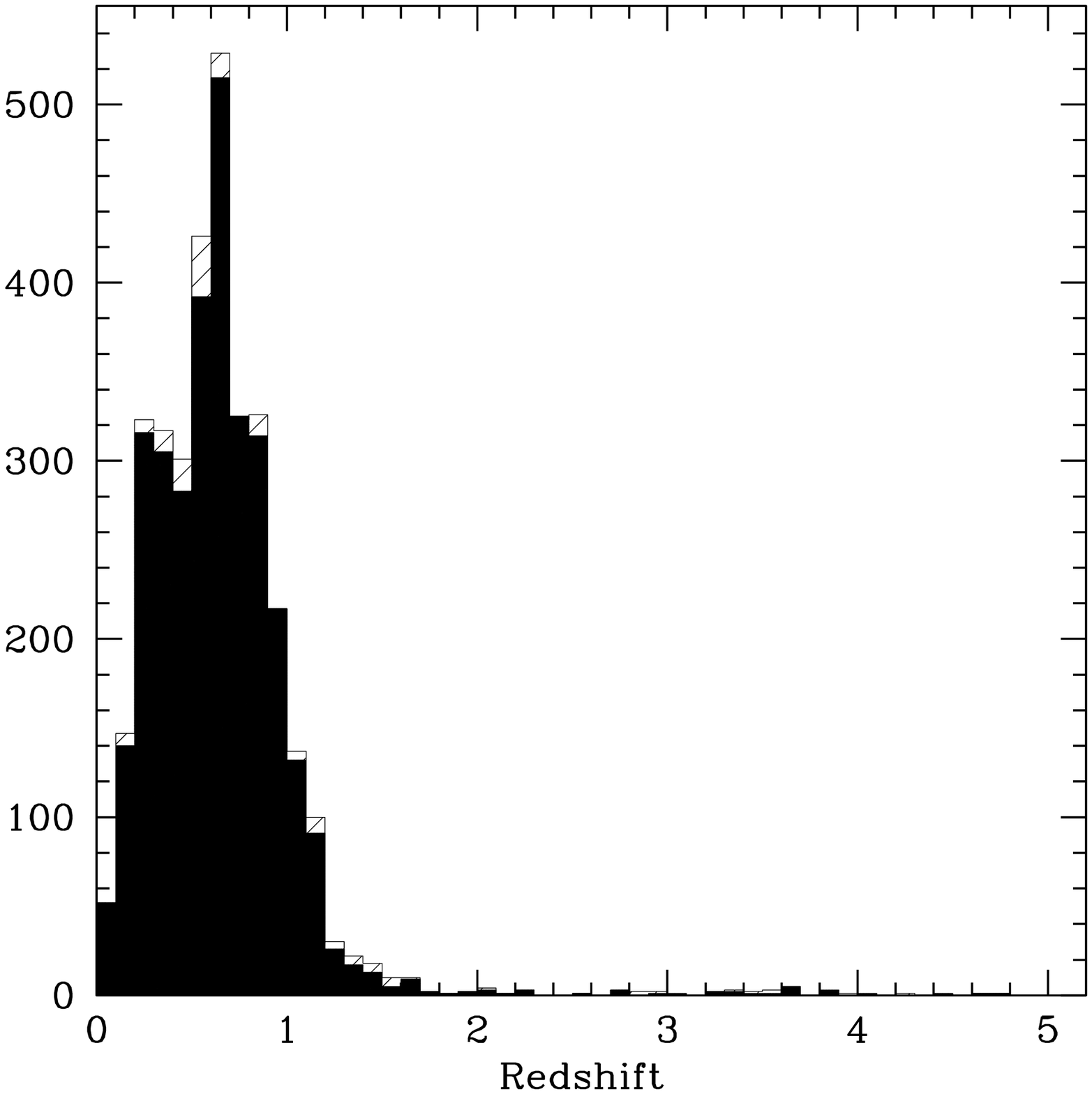}
      \caption{Redshift distribution for a magnitude-limited
sample of 3621 galaxies with $17.5 \leq I_{AB} \leq 22.5$, 
extracted from the sum of the VVDS-CDFS and the
VVDS-F02h fields. Galaxies with flags 2,3,4 and 9 are represented
by the filled histogram, galaxies with flag 1 by the open
histogram. This sample, including flags 2,3,4 and 9, 
is 93\% complete. The median
redshift is $z=0.62$ and the mean redshift is $z=0.65$.
              }
         \label{histz225}
   \end{figure*}

\begin{figure*}
   \centering
   \includegraphics[width=\textwidth]{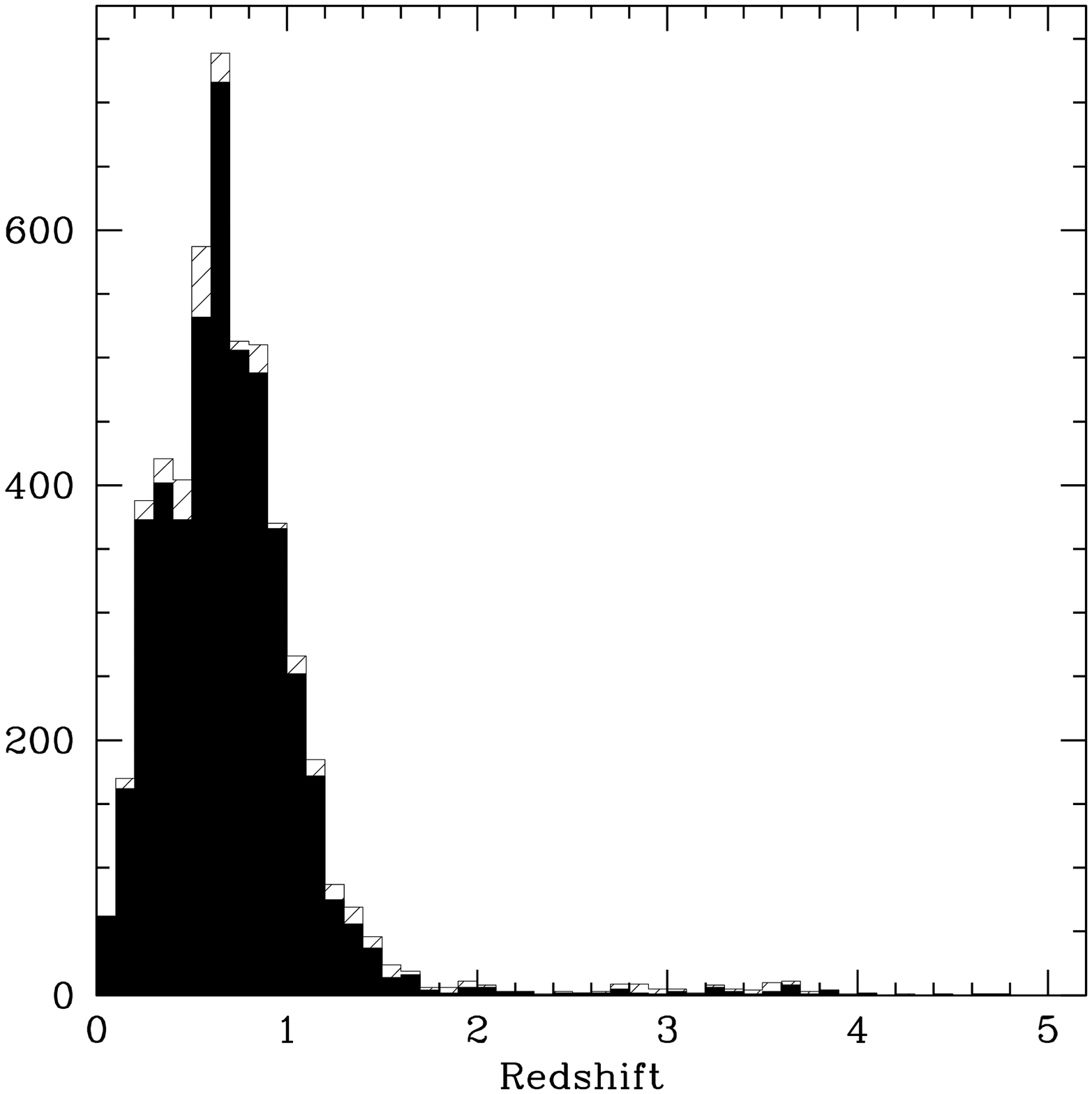}
      \caption{Same as Figure \ref{histz225} for 
5124 galaxies with $17.5 \leq I_{AB}\leq23$. 
This sample, including flags 2,3,4 and 9,  is 90\% complete. The median
redshift is $z=0.67$ and the mean redshift is $z=0.72$.
              }
         \label{histz23}
   \end{figure*}

\begin{figure*}
   \centering
   \includegraphics[width=\textwidth]{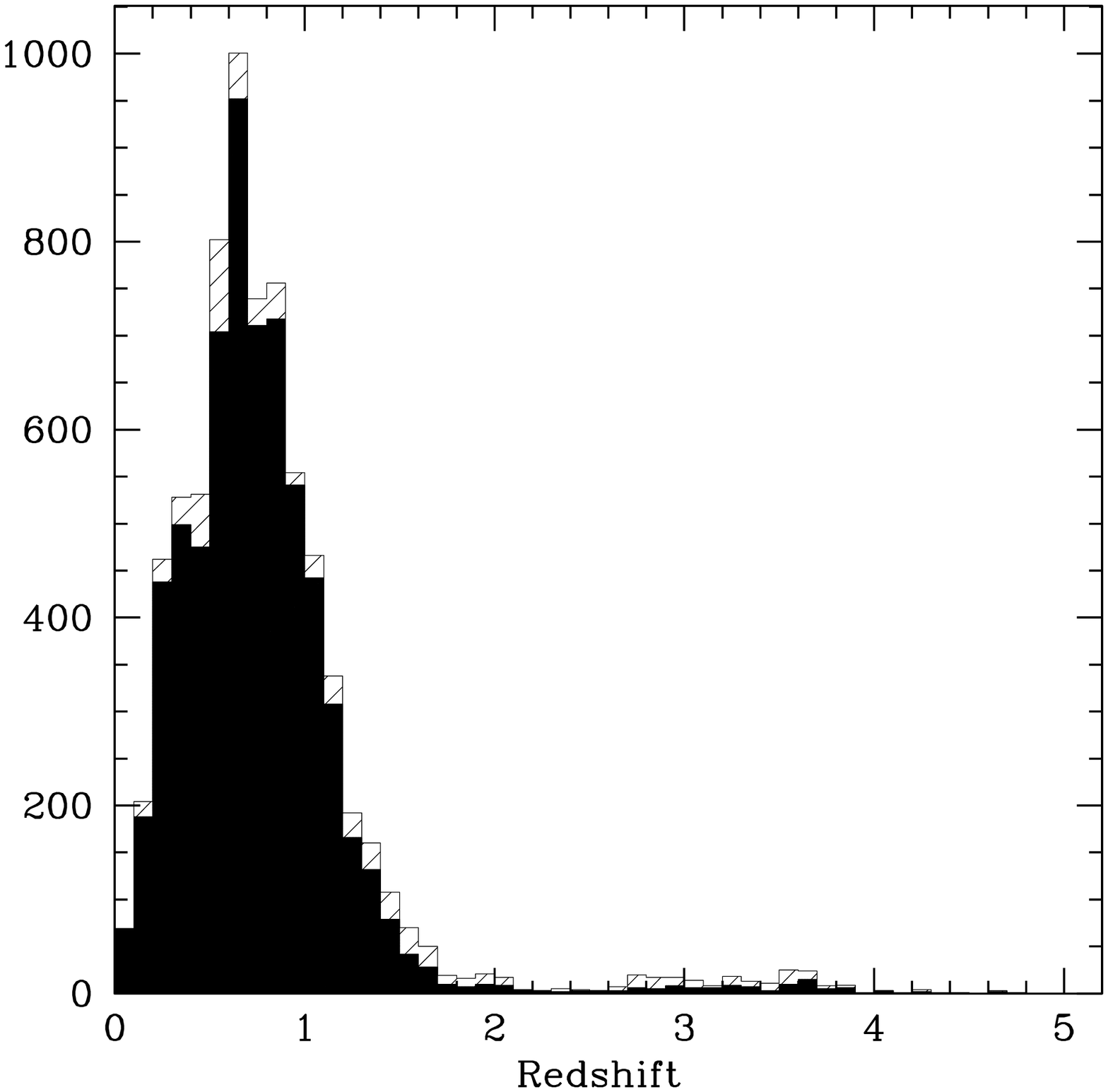}
      \caption{Same as Figure \ref{histz225} for  
7054 galaxies with $17.5 \leq I_{AB}\leq23.5$. 
This sample, including flags 2,3,4 and 9,  is 85\% complete. The median
redshift is $z=0.70$ and the mean redshift is $z=0.80$.
              }
         \label{histz235}
   \end{figure*}

From the $17.5 \leq I_{AB} \leq24$ sample, we can extract samples 
magnitude-limited down to $I_{AB}\leq22.5$,
$I_{AB}\leq23$, and $I_{AB}\leq23.5$, which are 
93\%, 91\%, 86\% complete respectively including only flags 2,3,4 and
9 (99\%, 98\%, 96\% including objects with flags 1). The redshift
distributions shown in Figures \ref{histz225} to
\ref{histz235} are therefore very secure. The redshift distribution
shifts slightly to higher redshifts going from a median redshift
of $z=0.62$ for $17.5 \leq I_{AB}\leq22.5$, to $z=0.70$ for
$17.5 \leq I_{AB}\leq23.5$. The median redshift and 1st and 3rd quartiles
are reported in Table ~\ref{zmed}. There is only a marginal high redshift
tail appearing at $z\geq2$ in the faintest magnitude bin to $I_{AB}=23.5$.
The redshift distribution for a sample down to $I_{AB}\leq23.5$ does
not go down significantly deeper in redshift than 
a sample limited to $I_{AB}\leq22.5$,
a fact to take into account when planning future redshift surveys.

\subsection{Redshift distribution for $17.5 \leq I_{AB}\leq24$}

\begin{figure*}
   \includegraphics[width=\textwidth]{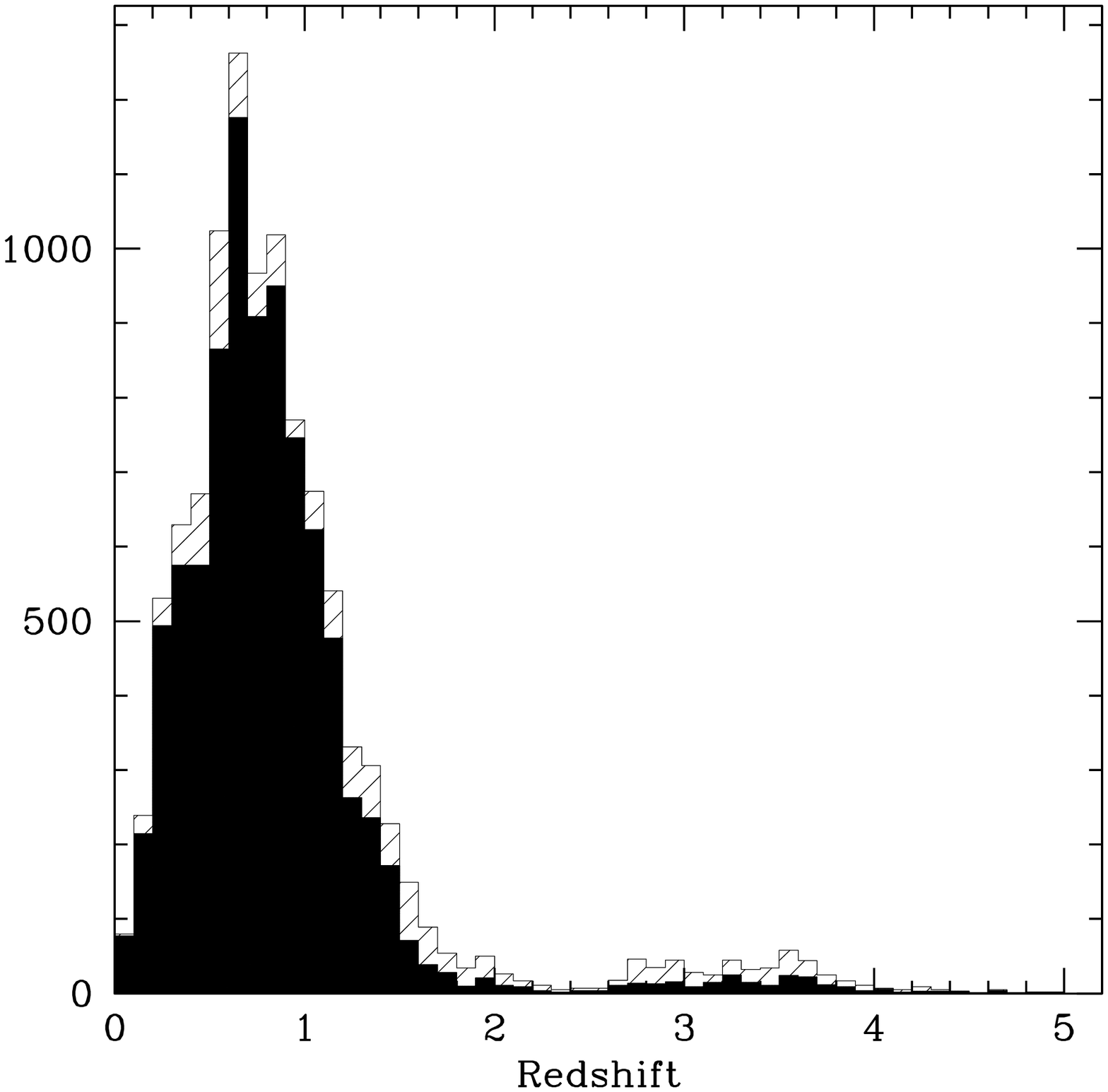}
      \caption{Same as Figure \ref{histz225} for  
9141 galaxies with $17.5 \leq I_{AB}\leq24$. 
This sample, including flags 2,3,4 and 9,  is 78\% complete. The median
redshift is $z=0.76$ and the mean redshift is $z=0.90$.
              }
         \label{histz24}
   \end{figure*}

The redshift distribution of the full First Epoch VVDS-Deep sample
is shown in Figure \ref{histz24}. A total of 9340 primary
target galaxies is present in this distribution including the VVDS-02h and 
VVDS-CDFS fields. The sample has a median redshift $z=0.76$,
and a significant high redshift tail appears up to 
$z=5.123$. The 
$17.5 \leq I_{AB} \leq 24$ magnitude selection allows us to 
continuously sample the galaxy population at all redshifts.
The VVDS spectroscopic 
measurement of the N(z) distribution improves upon previous
determinations using photometric redshifts (\cite{brodwin}),
in particular in the range $1.4 \leq z \leq 5$ where the
number of galaxies is small compared to the fraction of
catastrophic failures generally encountered when
computing photometric redshifts.  
This fraction of galaxies
appearing at all redshifts $2<z<\sim5$ indicates that
the observations are deep enough to probe the brightest part
of the population at these redshifts. 
The gap in the redshift distribution in the
interval $2.2 \leq z \leq 2.7$ is readily understood
in terms of a lower efficiency in measuring redshifts
for galaxies in this range because of the VVDS observed
wavelength range (see Section 6.2 and Paltani et al.,
in preparation). 
Objects with $z>2$ are examined in detail,
in particular to evaluate the fraction of 
flag 1 (and to a much lesser extent, flag 2) galaxies
lying at these redshifts; and the properties of this
population will be discussed elsewhere (Paltani et 
al., in preparation). 

   \begin{table*}
      \centering
      \caption[]{Median redshifts, 1st and 2nd quartiles of the redshift distribution
vs. magnitude (primary targets only)}
      \[
        \begin{array}{lcccc}
           \hline
            \noalign{\smallskip}
I_{AB} $range$ &  N_{obj} ($flags$ \geq2)  & 1^{st}$quartile$  & $Median$  & 3^{rd}$quartile$   \\
            \noalign{\smallskip}
            \hline
            \noalign{\smallskip}
18.0-18.5      &    9 & 0.13 &  0.18 &  0.24 \\ 
18.5-19.0      &   23 & 0.21 &  0.30 &  0.37 \\ 
19.0-19.5      &   45 & 0.21 &  0.30 &  0.41 \\ 
19.5-20.0      &   67 & 0.24 &  0.36 &  0.52 \\ 
20.0-20.5      &  111 & 0.26 &  0.43 &  0.64 \\ 
20.5-21.0      &  228 & 0.32 &  0.50 &  0.68 \\ 
21.0-21.5      &  361 & 0.38 &  0.58 &  0.76 \\ 
21.5-22.0      &  569 & 0.42 &  0.65 &  0.89 \\ 
22.0-22.5      &  833 & 0.46 &  0.70 &  0.91 \\ 
22.5-23.0      & 1156 & 0.50 &  0.79 &  1.11 \\ 
23.0-23.5      & 1493 & 0.56 &  0.85 &  1.17 \\ 
23.5-24.0      & 1675 & 0.58 &  0.90 &  1.32 \\ 
            \noalign{\smallskip}
            \hline
         \end{array}
      \]
\label{zmed}
   \end{table*}

\subsection{Redshift distribution in the VVDS-02h field}

\begin{figure*}
   \includegraphics[width=\textwidth]{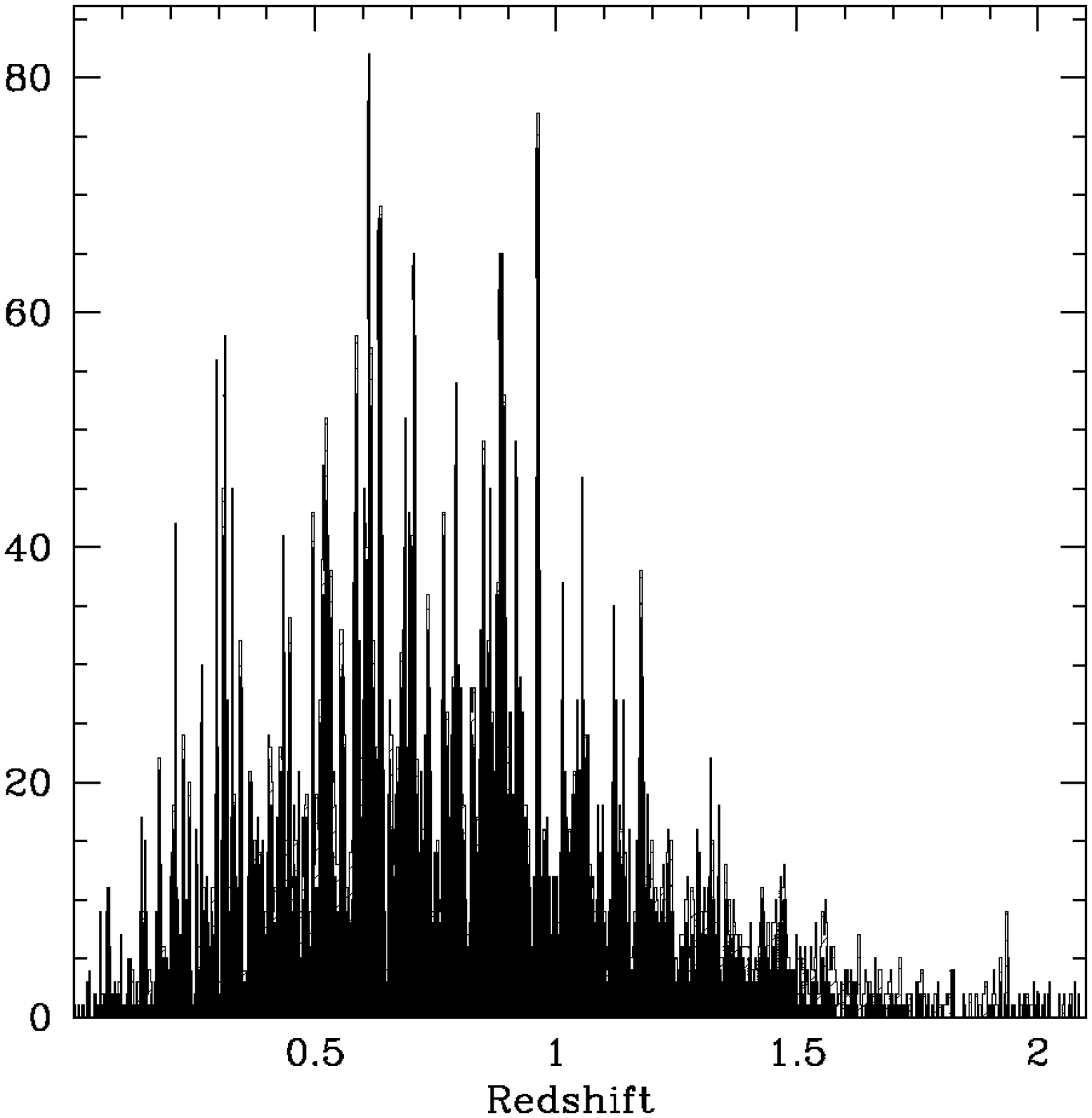}
      \caption{Redshift distribution for galaxies in the VVDS-02h
field with $0 \leq z \leq 1.7$, for
galaxies with redshift quality flags 2,3,4,9 (solid histogram)
and galaxies with redshift quality flag 1 (dashed histogram).
The redshift bin is dz=0.0033.
              }
         \label{histz02_1}
   \end{figure*}

The redshift distribution of galaxies in the VVDS-02h field
is presented in Figure \ref{histz02_1}
with smaller redshift bins.
The field shows an alternance of strong
density peaks and almost empty regions, with strong
peaks identified all across the redshift range, although
less prominent at $z>\sim2$. This is the first time
that the large-scale structure distribution of galaxies has been probed
on transverse scales $\sim30$h$^{-1}$ Mpc at these redshifts.
A detailed analysis of the clustering properties of
galaxies will be published elsewhere (\cite{olfcorr}, \cite{marinoni}).

\section{Conclusion and summary}

We have presented the strategy and the first epoch data of the 
VVDS, an I-band limited deep redshift survey of the distant
universe. We have been able to observe 11564 objects 
in the range $0 \leq z \leq 5.2$, in 
a total area of $\sim2200$arcmin$^2$ in the
VVDS-02h and VVDS-CDFS fields, which is thus the largest deep redshift
survey at a depth $17.5 \leq I_{AB} \leq 24$ to date.

The multi-slit observations with the VLT-VIMOS instrument
and heavy data processing with our VIPGI pipeline
are described. Emphasis has been placed on the measurement
of redshifts using the KBRED automated redshift measuring 
engine and the associated quality control. Independent 
redshift measurements have been produced and compared
between two survey members, a final check was performed
by a third party, and quality flags assigned to
each redshift measurement. Repeated observations of the same objects in
independent observations have allowed us to quantify the
velocity accuracy of the survey as $276$km/s. From these repeated
measurements,
and using photometric redshifts, we have been able to quantify 
the reliability of each of the redshift quality flags, 
and hence to derive the completeness
of the survey in terms of the fraction of secure redshift
measurements.

A total of 9340 redshifts have been measured on primary
targets (7840 objects have the most secure flags 2, 3, 4 and 9), 
and an additional 342 redshifts were obtained on secondary
targets falling in the multi-slits by chance (of which
260 are the most secure). A total of 603 galaxies (305 with
the most secure redshifts) have been measured in the
range $1.4 \leq z \leq 2.5$, and 462 in the range
$2.5 \leq z \leq 5.2$ (186 with the most secure redshifts).
Without a priori compactness selection in the photometric catalog
our spectroscopic sample includes 836 galactic 
stars, but 90 QSOs have been successfully identified. Following the
quality flags associated with the redshift measurements,
the sample is 78\% complete (secure redshifts), while 93\%
of the sample has a redshift measured. We have presented
the core properties of the sample in terms of spatial
distribution, absolute magnitude and B-I color vs. redshift,
and presented examples of observed spectra, 
revealing the wide range of galaxy types and luminosities
present in the survey.  

We also presented the redshift distribution
of magnitude limited samples down to $I_{AB}=24$.
For samples purely selected in I-band magnitude,
with $17.5 \leq I_{AB} \leq 22.5$,
$17.5 \leq I_{AB} \leq 23.0$ and 
$17.5 \leq I_{AB} \leq 23.5$, we find  a median redshift
of $z=0.62$, $0.67$ and $0.70$, respectively. For the complete
first epoch $17.5 \leq I_{AB} \leq 24.0$ VVDS-Deep sample,
we find that the median redshift is $z=0.76$, with a significant 
high redshift tail $1.5 < z < 5.2$ readily apparent.

The first epoch VVDS dataset presented here is used
extensively by the VVDS team to measure evolution in
the galaxy population, as presented in joint papers and
several papers in preparation. It provides an unprecedented
sample to study galaxy evolution over 90\% of the
life of the universe.

\begin{acknowledgements}
This research has been developed within the framework of the VVDS consortium
(formerly the VIRMOS consortium).\\
This work has been partially supported by the 
CNRS-INSU and its Programme National de Cosmologie (France)
and by Italian Research Ministry (MIUR) grants
COFIN2000 (MM02037133) and COFIN2003 (num.2003020150).\\
The VLT-VIMOS observations have been carried out on guaranteed 
time (GTO) allocated by the European Southern Observatory (ESO)
to the VIRMOS consortium, under a contractual agreement between the 
Centre National de la Recherche Scientifique of France, heading
a consortium of French and Italian institutes, and ESO,
to design, manufacture and test the VIMOS instrument.
\end{acknowledgements}

\end{document}